\documentclass[journal, 12pt, onecolumn, draftcls]{IEEEtran}

\usepackage{cite}
\usepackage{amsmath}
\usepackage{amssymb,latexsym}
\usepackage{booktabs}
\usepackage{graphicx}
\usepackage{psfrag}
\usepackage{epsfig}
\usepackage[caption=false,font=footnotesize]{subfig}
\usepackage{setspace}
\usepackage{array}
\usepackage{mathtools}
\usepackage{units}
\usepackage{siunitx}
\usepackage[dvipsnames]{xcolor}
\usepackage{comment}
\usepackage{siunitx}

\newcommand{\dB}{\mathrm{dB}}
\newcommand{\degree}{\ensuremath{^\circ}}

\title{Achieving Phase Coherency and Gain Stability in Active Antenna Arrays for Sub-6 GHz FDD and TDD FD-MIMO: Challenges and Solutions}

\author{Reza Monir Vaghefi, Ramesh Chembil Palat, Giovanni Marzin, \\ Kiran Basavaraju, Yiping Feng, and Mihai Banu 
\thanks{Authors are with Blue Danube System Inc., Santa Clara, CA 95054. E-mail: \{vaghefi, rameshcp, marzin, kiranb, ypfeng, mihaibanu\}@bluedanube.com}
}

\begin{document}
\pagestyle{empty}

This paper was submitted for publication in IEEE Journal on Selected Areas in Communications
on August 29, 2019.

\textsuperscript{\textcopyright} 2019 IEEE. Personal use of this material is permitted. Permission from IEEE must be obtained for all other uses, in any current or future media, including reprinting/republishing this material for advertising or promotional purposes, creating new collective works, for resale or redistribution to servers or lists, or reuse of any copyrighted component of this work in other works.

\maketitle
\thispagestyle{plain}
\pagestyle{plain}

\vspace{-2cm}
\begin{abstract}
Massive MIMO has been the subject of intense interest in both academia and industry for the past few years. 3GPP standardization for cellular systems have adopted the principles of massive MIMO and categorized the use of large rectangular planar arrays at the base station as full-dimension MIMO (FD-MIMO) to operate in both TDD and FDD. Operating a large antenna array base station requires the system to overcome several implementation challenges caused by hardware impairments making practical solutions non-ideal and expensive to deploy at scale. It is important to learn from existing challenges and solutions in order to prepare for larger scale deployment for example with cell free massive MIMO. Hence in this paper, we specifically study the phase and amplitude instability due to RF impairments using measurements carried out in the lab and in the field in a commercial LTE network. We investigate the effect of phase and magnitude errors on the performance of FD-MIMO systems. We discuss and characterize various sources creating these errors including time varying phase drift from low cost local oscillator (LO) and internal temperature variations affecting frequency response of RF chains. The minimum requirements and tradeoffs of different LO architectures and calibration mechanisms for practical cellular deployment are discussed. We then provide details of a novel coherent LO distribution mechanism and related novel array calibration mechanism that can be applied to both TDD and FDD systems. Measurement results are provided to validate the performance of these methods used in a 2D full-connected hybrid beamforming array architecture called High Definition Active Antenna System (HDAAS). These results showcase the efficacy of the proposed methods which can easily be extended to other array architectures including sub-array hybrid beamforming and element-level digitization. 
\end{abstract}

\begin{IEEEkeywords}
Massive MIMO, FD-MIMO, calibration, phase and magnitude error, array coherency
\end{IEEEkeywords}

\section{Introduction}
Massive multiple-input and multiple-output (MIMO) using large 2D planar active antenna arrays is currently viewed as a key solution for solving capacity demands in cellular networks for sub-6 GHz bands such as Long-Term Evolution (LTE) Advanced Pro and fifth generation (5G). Massive MIMO is classified as full-dimension MIMO (FD-MIMO) in 3rd Generation Partnership Project (3GPP) documents~\cite{SamsungFDmimo2017}. It refers to the use of a large number of antennas at the base station (BS) to support simultaneous transmission to multiple user equipment (UE) terminals (referred to as multi-user MIMO or MU-MIMO) to achieve significant improvement in network throughput. It is a topic of active research in academia and industry for 3GPP 5G mobile networks standardization~\cite{ShafiJsac20175G}.

\subsection{FD-MIMO in TDD and FDD Systems} \label{FdmimoTddFdd}

A good overview of FD-MIMO system and the terminology used to describe different active antenna architectures can be found in~\cite{SamsungFDmimo2017,3gppAAs2015} and similar terminology will be used in this paper. Here we briefly review relevant background related to FD-MIMO.

FD-MIMO supports both channel reciprocity and channel feedback mechanisms to determine the precoders for downlink transmission. Channel reciprocity is the predominant method used to determine the precoder in TDD systems. If uplink and downlink radio frequency (RF) chains are well matched or calibrated, the channel state information (CSI) determined in uplink can be used to compute the precoder using any one of the methods found in literature \cite{yang2013performance}. Since interference is not reciprocal in cellular deployment~\cite{Ericsson2016, Nokia3GPP2017} and uplink channel estimation errors can become worse especially for cell edge users, 3GPP proposes using explicit channel feedback mechanism for TDD FD-MIMO. Channel feedback mechanism is default for FDD systems where reciprocity cannot be applied. Two basic CSI feedback mechanisms are proposed in 3GPP for FD-MIMO~\cite{SamsungFDmimo2017,Vook2018}; non-precoded codebook and precoded or pre-beamformed channel sounding using CSI-reference signals (CSI-RS). In the non-precoded codebook feedback mechanism, the UE selects the best index from a codebook which can then be used to determine the downlink precoder for data transmission. In the beamformed CSI-RS approach, a grid-of-beams, each with a distinct CSI-RS signal, is transmitted by BS. Then, the UE reports back either the best beam or a linear combination of beams to the BS~\cite{Ericsson5GNRTestbed2018}. 

FD-MIMO accommodates a combination of analog precoder and digital precoder for beamforming data streams. Specific combination depends on the mapping of radiating elements to digital streams or antenna ports. The antenna array configuration is represented by (M, N, P), where M is the number of rows, N is the number of columns, and P is the number polarizations. Port configuration is represented by (V, H), where V and H are the number of ports in vertical and horizontal direction, respectively. An example antenna array with (12,4,2) configuration consists of 12 rows and 4 columns with 48 dual-polarized antennas in total. Consider the case of (4,4) configuration where the array combines 3 elements of a column to form an antenna port where they share a common data converter. This means from the digital side each column gets configured with 4 antenna ports per polarization. Hence, the array would then consist of 32 antenna ports in total. Note that in some cases analog beamforming weights can still be used to create a specific radiation pattern for each 3-element sub-array. Analog beamformer can be viewed as a time domain beamformer. Traditionally, these are implemented using analog phase shifters. However, in recent years novel array implementations using mixed signal integrated circuit (ICs) that provide digital phase and gain control in time domain have become practical~\cite{Vaghefi2018, Anokiwave2017}. These ICs can replace the analog phase shifters and update the time domain beam patterns at a much faster rate. Unlike analog beamforming, the digital precoder part can be applied in both frequency and time domain per time/frequency resource unit. Hence, element-level digitization supporting only a digital precoder, hybrid beamforming with sub-array architecture and hybrid beamforming with 1D or 2D full-connection architecture are all possible configurations to implement FD-MIMO~\cite{3gppAAs2015}. In the case of codebook-based feedback, the CSI-RS signals are each mapped to a portion of the array using digitized antenna ports which is well suited for element-level digitized array or sub-array hybrid beamforming architectures~\cite{SamsungFDmimo2017}. In the case of beamformed CSI-RS feedback, both sub-array and full-connected hybrid beamforming architectures can be employed~\cite{SamsungFDmimo2017}. Each CSI-RS symbol associated with a CSI-process is mapped to a pre-determined beam pattern. The proper choice among the proposed architectures for commercial deployment depends on tradeoff between performance, design complexity, and cost. This leads to the next discussion on challenges with hardware impairments that is critical to understand in making these tradeoffs.

\subsection{Challenges in Field Operation Due to Hardware Impairments}
Large antenna installation over tower tops and buildings for a macro-cell network are severely limited in terms of space and weight due to wind loading constraint. Building and operating a large active antenna array is made challenging due to the following factors: design complexity from large number of high speed digital I/O and corporate feeds; supporting large digital processing section required to implement digital front end (DFE) functions and lower-level physical layer processing functions; high power consumed by the analog front end (AFE) components, data converters and the challenges with thermal design for heat dissipation within a sealed enclosure to meet weight constraints~\cite{Honcharenko2019}. The RF conformance requirements in the commercial licensed bands are also very stringent in terms of adjacent channel leakage ratio (ACLR) and error vector magnitude (EVM)~\cite{3GPP2017104, 3GPP2019} limiting simplification of the AFE and DFE designs in terms of data converter resolution and PA efficiency. In addition to the above the design complexity also involves resources utilized to overcome non-ideal behavior due to hardware impairments.

Theoretical aspects of massive MIMO have been widely studied in the literature. Proof of concept testbeds in academia such as~\cite{Zhong2012,Tufvesson2014,HausteinPoCMamimo2016, Beach2018} and some field trials under controlled test environments by the industry such as~\cite{Wei2017,Kishiyama2017,Ericsson5GNRTestbed2018} have conducted feasibility studies in the past. However, there is not much experimental data available characterizing and validating solutions that overcome hardware impairments and corresponding impact on field performance in real world network deployment. Very limited work, for example~\cite{Okumura2017,Okumura2018}, have showcased the measured performance that is impacted by hardware impairments causing channel reciprocity errors. Even in these experiments, the active antenna system is operated under very low transmit power ranging from 25-37 dBm compared to a typical macro-cell requiring 46-52 dBm, which do not really stress the hardware design due to heat dissipation. 

The effects of hardware impairments are broadly categorized into residual additive noise, multiplicative noise and thermal noise~\cite{Debbah2015,Ratnarajah2017}. Residual additive noise is attributed to the combination of non-linear behavior of AFE (e.g., signal clipping from power amplifiers), quantization noise from the data converters and IQ imbalance. Multiplicative noise is attributed to phase noise from LOs and from phase and amplitude variations between RF chains due to variations in frequency response of analog components. Solutions such as crest factor reduction, digital pre-distortion, high resolution data converters and IQ imbalance correction are commercially available to mitigate the effect of residual additive noise. However, impairments that result in multiplicative noise are harder to solve and require custom solutions. Prior studies investigating methods to minimize multiplicative noise impairments have mostly focused around reciprocity based calibration~\cite{Kaltenberger2018}. However, there is limited experimental verification validating the efficacy of these methods on practical antenna array systems. For example~\cite{SamsungJsac2017FDMimoProto} showed calibration measurements with an active antenna system operating at low transmit power under controlled lab environment. It is unclear if this method requires blanking of active transmission during transmission and how effective it is under internal temperature variations. In~\cite{Tufvesson2017}, an over-the-air mutual coupling-based method was experimentally verified. However, over-the-air methods interfere with receive paths and hence such a method may not be applicable in practical cellular systems. Another aspect, even though there are many papers that have examined the issue~\cite{Larsson2015, Krishnan2016, PuglielliBwrcPnBF2016, Tufvesson2017, Dai2015, Luo2016, Tafazolli2017},  is the limited information connecting the desired levels of phase and magnitude errors to what is practically feasible with respect to actual components and practical design. There is a lack of experimental results validating synchronization or phase coherency of RF paths under operating conditions expected in a cellular deployment.

It is evident from the above discussion that evaluation of hardware impairments and efficacy of their compensation through experimental validation is still essential to understand practical feasibility and limitations for FD-MIMO systems. This will also help assess the challenges with active antenna solutions to scale to larger array sizes for future use cases. In this paper, we address practical issues and describe field tested solutions to the problem of multiplicative noise.

\subsection{Summary of Results}
This paper includes the following contributions:
\begin{itemize}
    \item Link and system level simulations investigating the impact and limits of phase and amplitude errors on both FDD and TDD FD-MIMO in multi-cell scenario,
    
    \item Analyzing the effects of short term LO phase drifts on beamforming in FD-MIMO, 
    
    \item Comparison of different LO distribution schemes given loss of coherency from phase drifts,
    
    \item Lab and field characterization of the temperature effects on phase and amplitude in a large array, deriving requirements for a good calibration method,
    
    \item Introducing a new, scalable, field-validated solution to low-cost FDD and TDD FD-MIMO with coherent LO distribution and accurate on-line calibration requiring no data-flow interruptions. 
    
\end{itemize}

\section{Effect of Phase and Amplitude Errors on FD-MIMO Performance} \label{phAmpErPerf}
Prior work have mostly analyzed effects of residual phase and amplitude errors for channel reciprocity based TDD massive MIMO systems in single cell scenario. Extending the analysis to multi-cell scenario along with pilot contamination, inter-cell interference, overhead due to channel feedback is non trivial. Hence in this section, we resort to Monte-Carlo simulations to compare the average sum-throughput in a multi-cell FD-MIMO system to understand the net impact of phase and amplitude errors from hardware impairments. A summary of the assumptions made for the simulation is provided in Table~\ref{tab:para}. The simulation is divided into two levels: the system level and the link level. In the system level simulation, a regular hexagonal cellular network with 19 sites and 57 cells is generated. 10 users per cell are randomly placed in a 3D area covered by the cells. Large scale and small scale channel parameters between each cell and user are generated using 3D channel model in~\cite{3gpp:3dchannel}. Channel large scale parameters are used to determine the received signal strength from each cell to each user. The user received power is used to determine the serving cell for each user. Channel small scale parameters are generated between each antenna element of the base station and each antenna element of the user. In the link level simulations, waveforms similar to LTE downlink and uplink are generated and passed through the 3D channel defined by the small scale parameters. The throughput is then calculated based on the received signal from the serving cell and interfering signals from neighboring cell at the user.

The antenna configuration used for the simulations is (12,4,2). The carrier frequency is 2 GHz and antenna element spacing is $0.5\lambda$. This specific antenna configuration is chosen for simulation since it falls within acceptable size limits imposed on commercial tower tops. The bandwidth of the system is 10 MHz in FDD and 20 MHz in TDD (with 50\% time allocated to downlink). The base station antenna element gain is 5 dBi. UEs have two antenna elements with a gain of 0 dBi. Users are scheduled in a round robin fashion and a full buffer traffic is considered. UEs use CSI-RS in LTE downlink waveform to report precoding matrix indicator (PMI) and channel quality indicator (CQI) and use demodulation reference signal (DMRS) for coherent reception in the downlink over precoded data as described in 3GPP for FD-MIMO~\cite{SamsungFDmimo2017}. CQI/PMI reporting periodicity is 5 ms and feedback delay is 6 ms. In the TDD system, the base station estimates the uplink channel using sounding reference signals (SRS) and applies zero forcing (ZF) or matched filter (MF) methods to determine the downlink precoder. Perfect channel estimation is assumed since this isolates the effect of hardware imperfections on performance.  In the simulated FDD system case, the digital precoder is obtained from the PMI reported by the UEs. Non-precoded CSI-RS with codebook based wide band PMI feedback is considered in this case~\cite{SamsungFDmimo2017}. 

\begin{table*}[t!]
\footnotesize
\centering
\caption{Summary of simulation parameters.}
\begin{tabular} {  p{2.5cm}  p{4.5cm}  p{3cm} p{4cm}} \toprule
System Level 	& Assumption						& Link Level 		& Assumption\\ \midrule
Duplex method			& FDD and TDD						& Downlink scheduler 		& Round robin \\
System Bandwidth 		& 10 MHz (FDD), 20 MHz (TDD)		& Traffic model				& Full buffer \\
Center Frequency 		& 2 GHz 							& DL channel estimation		& Ideal on CSI-RS and DM-RS\\
Cellular layout			& 19 sites, 57 cells 				& Downlink receiver			& MMSE		\\
Network layout			& 3D hexagonal grid, wraparound 	& Interference modeling		& Explicit intercell interference\\ 
Inter-Site distance 	& 500 m								& Link adaptation			& MCS based on reported CQI\\
Users per cell			& 10 (uniformly in 3D space) 		& CSI (FDD)					& Class A non-precoded PMI \\
BS Antenna 				& (12, 4, 2)						& CSI (TDD) 				& Uplink SRS \\
BS element gain			& 5 dBi 							& Max number of layers  	& 4 (max 1 layer per user)\\ 
User noise figure		& 9 dB 								& Feedback delay			& 6 ms\\
\bottomrule
\end{tabular}
\label{tab:para}
\end{table*}

The performance of the FD-MIMO system is evaluated against phase and amplitude errors along with different number of digitized antenna ports. It is also assumed that even under phase errors the UEs are able to lock on to the carrier frequency and perform carrier frequency offset correction and synchronization using broadcast channels as we only evaluate the impact on the data channel. The number of digitized antenna ports considered in the simulations are 4, 8, 16, and 32 with (2,1), (2,2), (4,2), and (4,4) port configurations~\cite{Vook2018}. 

\begin{figure*}[t!]
\centering
\subfloat[FDD - Phase]{\label{fig:phase:fdd:1}\includegraphics[width=0.48\textwidth]{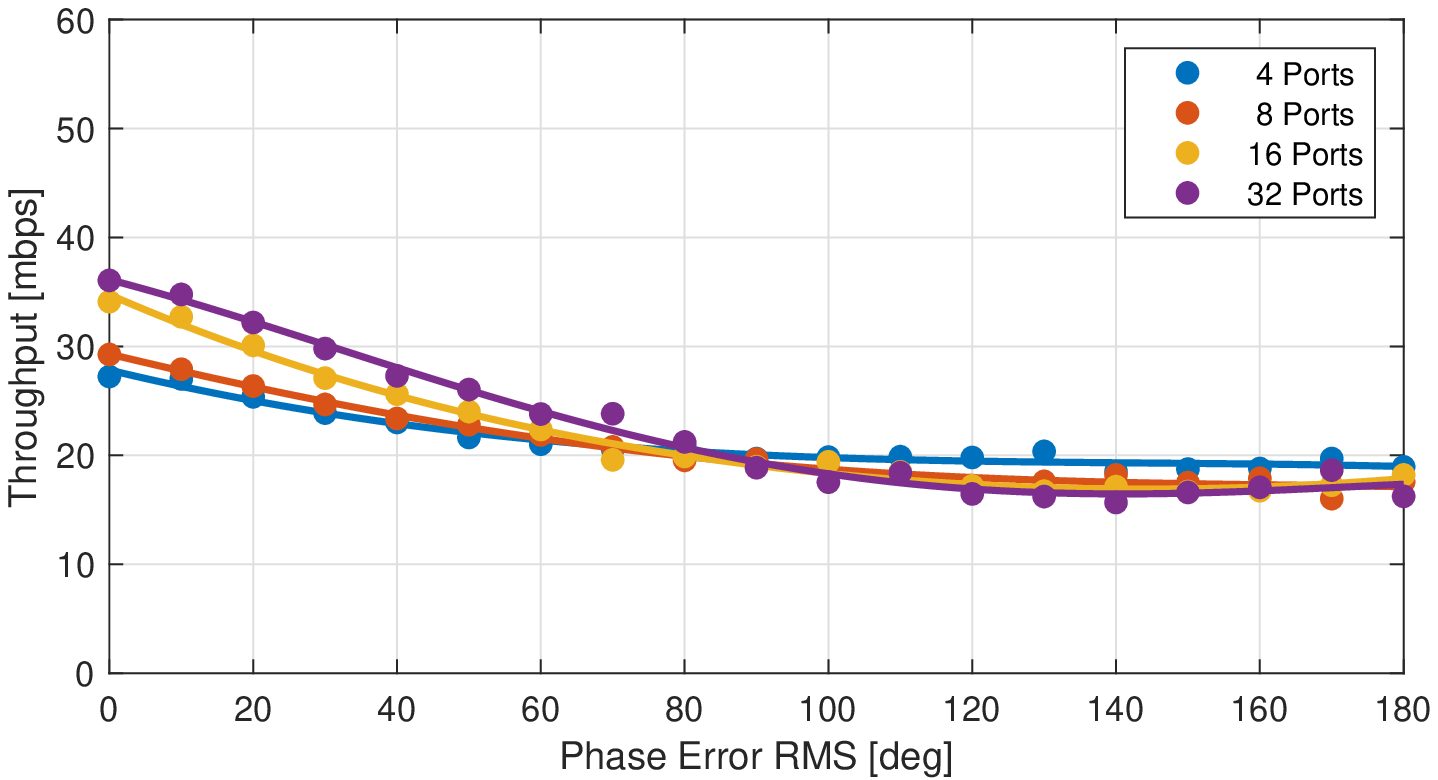}} \hfill
\subfloat[TDD - Phase]{\label{fig:phase:tdd:1}\includegraphics[width=0.48\textwidth]{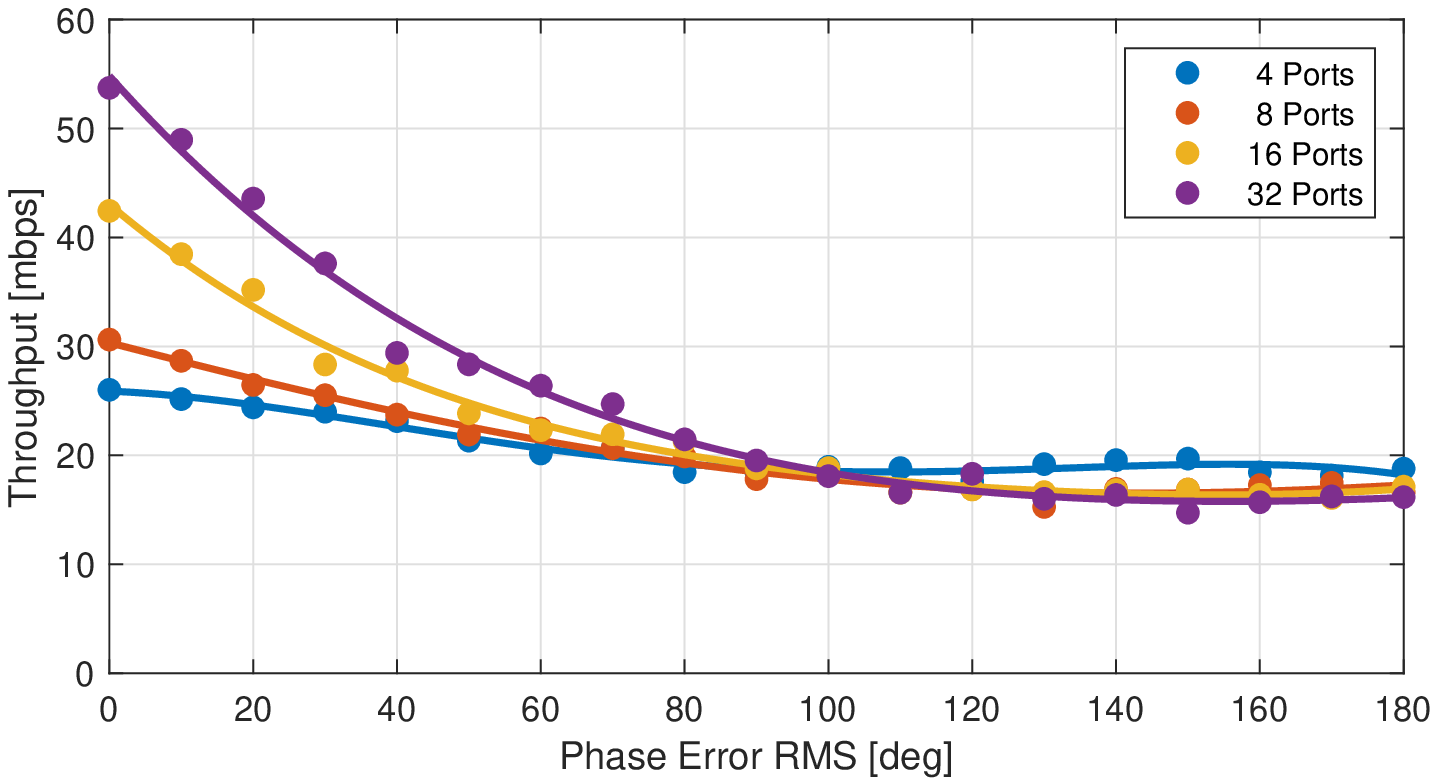}} \\
\subfloat[FDD - Magnitude]{\label{fig:mag:fdd:1}\includegraphics[width=0.48\textwidth]{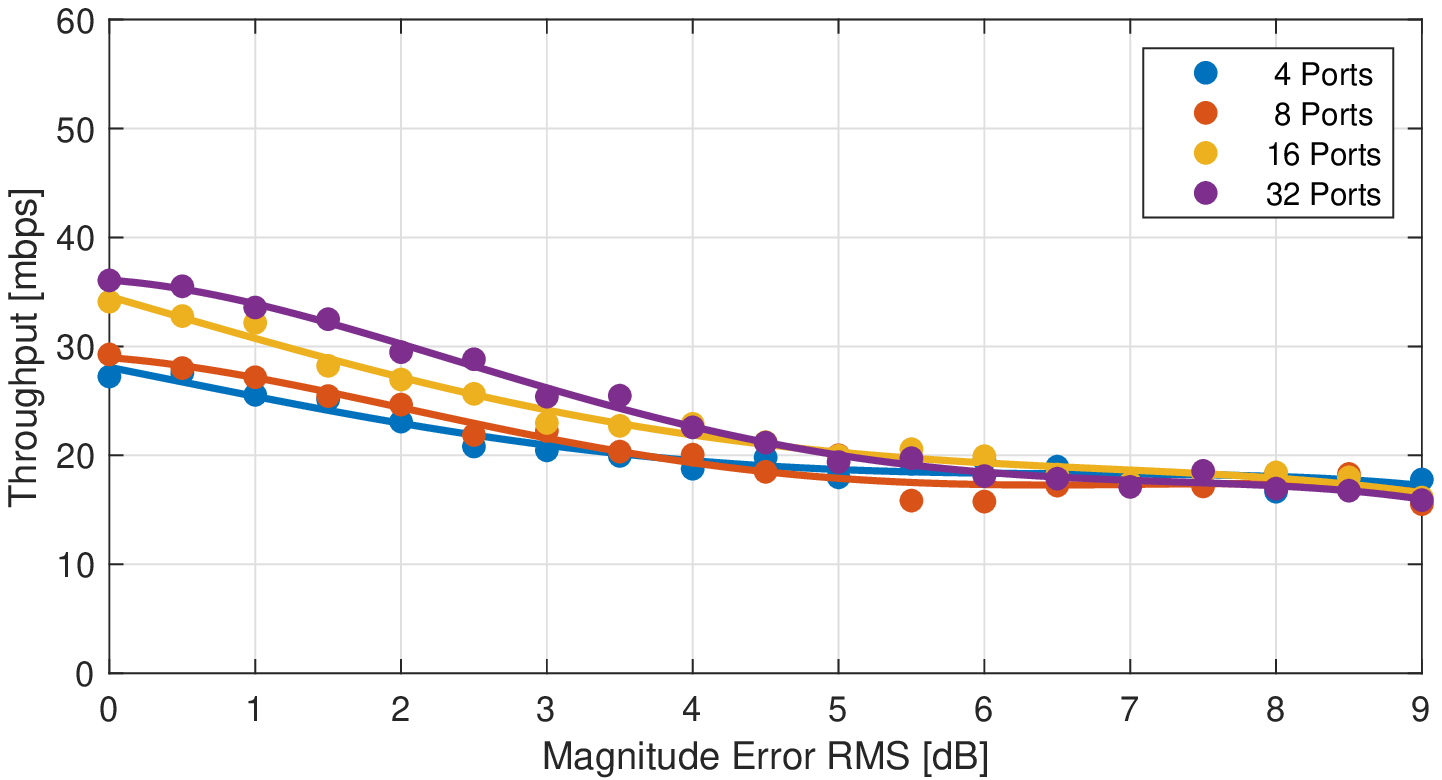}} \hfill
\subfloat[TDD - Magnitude]{\label{fig:mag:tdd:1}\includegraphics[width=0.48\textwidth]{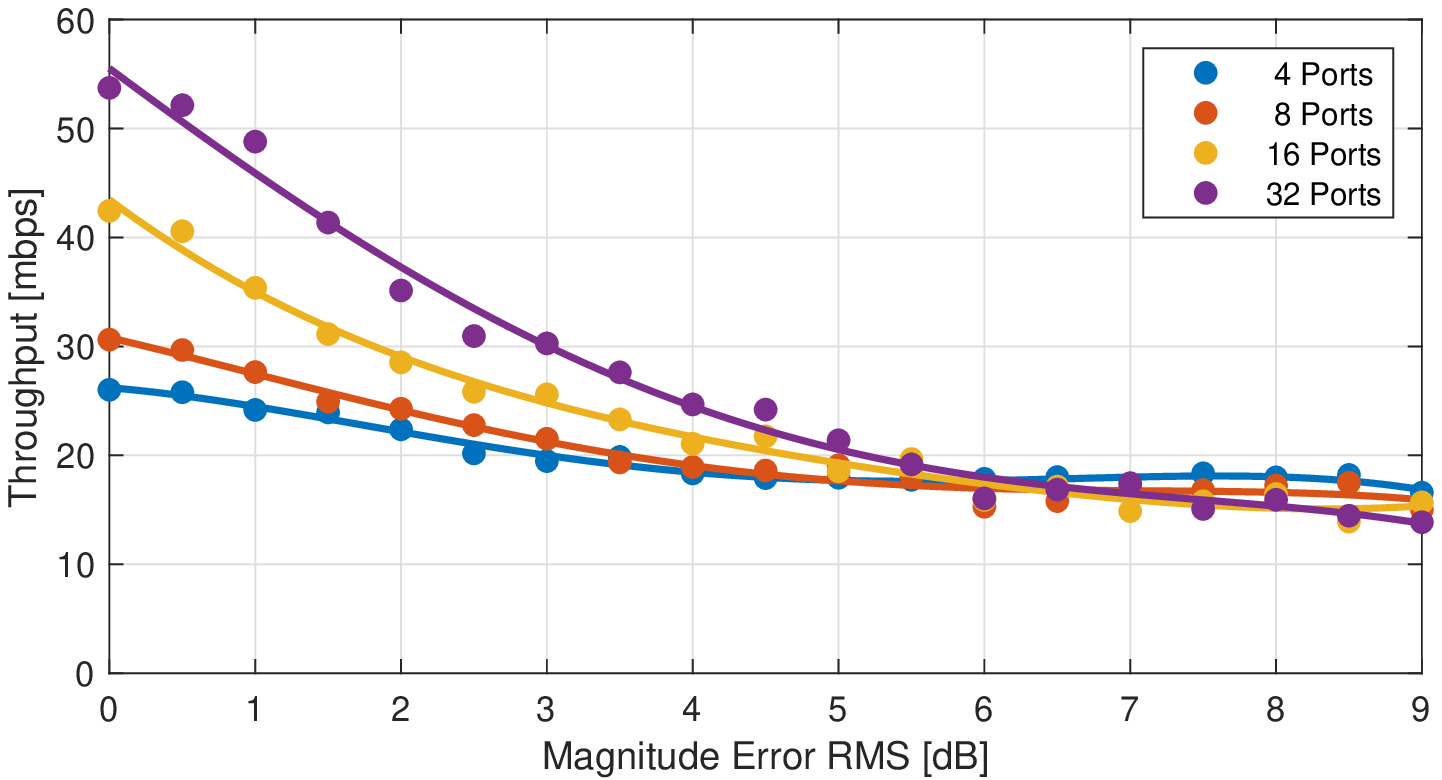}}
\caption{Throughput of a FDD FD-MIMO system under phase and magnitude calibration error.}
\label{fig:error}
\end{figure*}

The sum-throughput performance of the FDD FD-MIMO system as a function of root mean square (RMS) of the phase error and the number of digital ports is illustrated in Fig.~\ref{fig:phase:fdd:1}. We define phase error as the deviation in phase of the signal from that of the correct phase corresponding to chosen precoder being transmitted from a particular RF chain corresponding to an antenna port . The phase error is modeled as a zero-mean Gaussian distribution with standard deviation equal to the target RMS phase error. The system also implements a dynamic switch between SU-MIMO and MU-MIMO transmission where up to 4 UEs can be spatially multiplexed. When conditions do not favor improved performance compared to MU-MIMO the system falls back to SU-MIMO. Under ideal scenario without any phase error ($0\degree$ phase error), the sum throughput improves as the number of digitized antenna ports increases. Increasing the ports from 4  logically configured as (2,1) to 8 ports (2,2) does not improve the throughput significantly, since additional ports are added in the vertical domain and the resulting higher resolution in the digital precoder is not good enough to separate users in elevation. With 16 ports (4,2) where the digital precoder has more resolution in the azimuth with 1 port per column, the throughput improves by 20\% in comparison over 8 ports. At 32 ports, the digital precoder has better resolution in elevation supporting 4 ports in a column than 2 ports and the throughput improves by 25\% over a 8 ports system. This incremental gain from additional vertical ports suggests that it is better to add more ports in the Horizontal plane. Fig.~\ref{fig:phase:fdd:1} also shows that as the RMS phase error increases, the throughput decreases in all cases due to increase in inter-user interference. After a $80\degree$ RMS phase error the performance in all cases hit a floor around which the system chooses SU-MIMO. The 32 ports system starts performing below the 8 ports system with no phase errors around $50\degree$ phase error and below the 16 ports system at \(20\degree\). 

If peak phase error is assumed to be \(4\sigma_{\theta}\). Then at $50\degree$ RMS phase error which is $200\degree$ peak corresponds to a maximum difference of \(0.56\) wavelengths or about 278 ps in time alignment. At \(20\degree\) RMS this is equal to \(0.22\) wavelengths or 111 ps in time alignment. This suggests coherent systems with fewer antenna ports may provide better performance at lower cost compared to a system with higher number of ports that have a phase error greater than \(20\degree\) RMS.

Fig.~\ref{fig:phase:tdd:1} shows the average sum-throughput performance of TDD FD-MIMO system with ZF precoder as a function of RMS phase error and number of antenna ports. Similar to the FDD case, use of higher number of ports shows better performance. Compared to the FDD case, the TDD system shows better performance due to the use of channel reciprocity that provides better CSI than a quantized codebook-based feedback. However, as the phase error increases, the performance of TDD system degrades rapidly and approaches closer to the FDD system with greater than \(20\degree\) RMS phase error. The beamforming and spatial multiplexing gains degrade with increasing phase errors and the system switches fully to SU-MIMO around $80\degree$ RMS. This shows that the reciprocity-based systems provide higher performance only at very high coherency.  

\begin{figure}[t!]
\centering
\begin{minipage}[b]{0.48\textwidth}
\includegraphics[width=\textwidth]{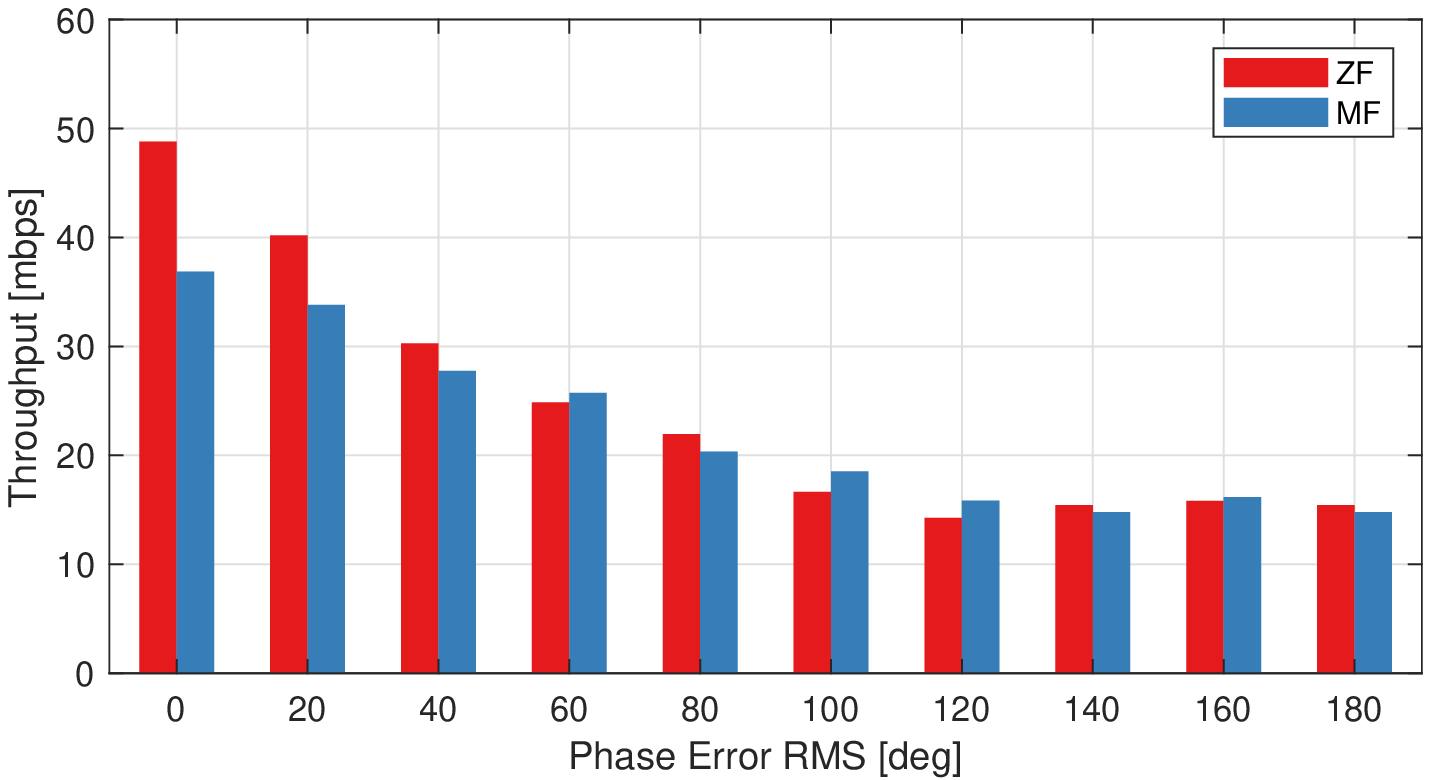}
\caption{Comparing ZF and MF digital precoder with 32 ports}
\label{fig:thr:mf} 
\end{minipage} \hfill
\begin{minipage}[b]{0.48\textwidth}
\includegraphics[width=\textwidth]{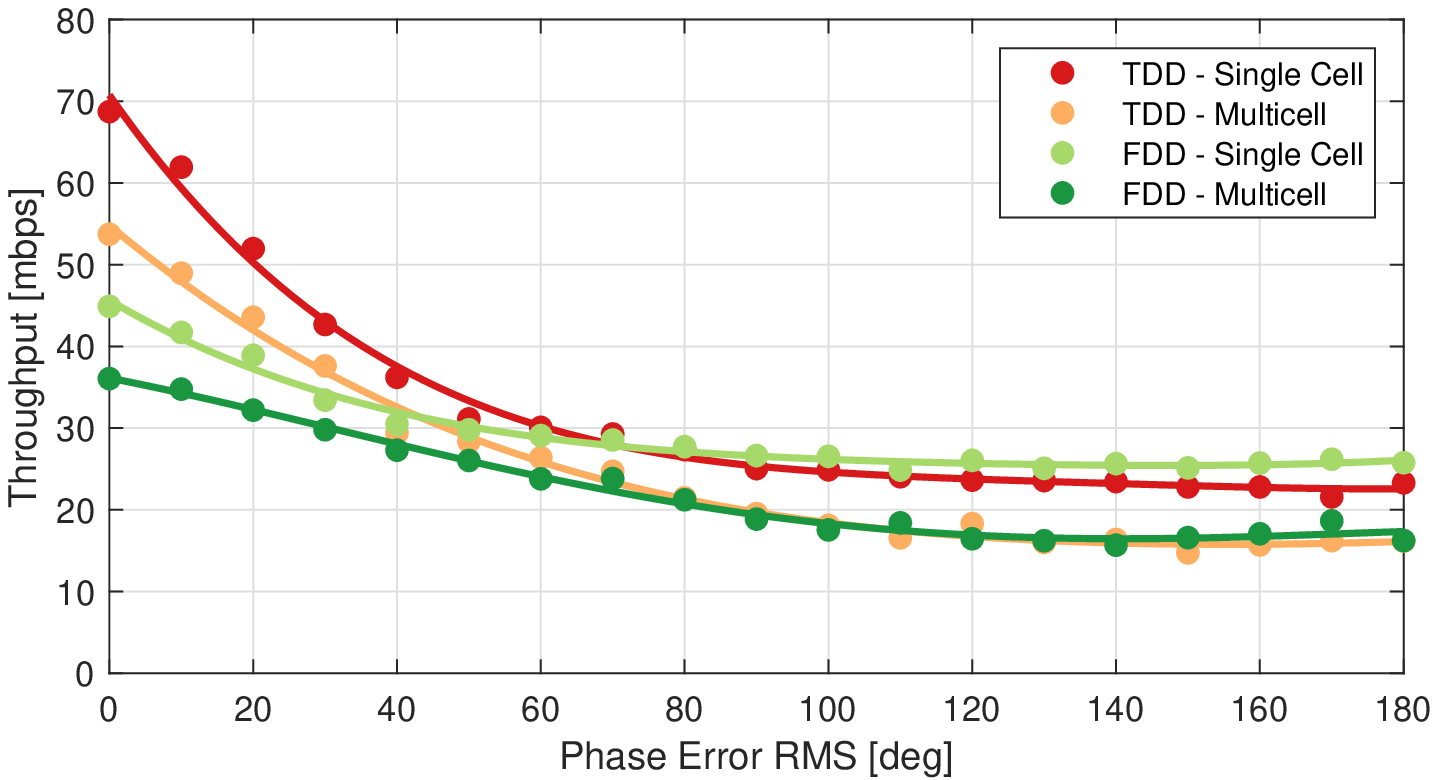}
\caption{Comparing single-cell and multi-cell scenarios}
\label{fig:single:multi} 
\end{minipage}
\end{figure}

One of the main reasons for the rapid degradation in TDD is the sensitivity of the nulls created by the beam pattern to phase errors. Analysis in~\cite{Madhow2012} shows that an ideal null depth of -25.62 dB from the peak power at the receiver locations of co-scheduled users with ZF precoder requires the transmit nodes to be within \(3\degree\) RMS phase error. Using same analysis, the null depth at \(20\degree\) RMS phase error decreases to -9.14 dB. This is a 15 dB increase in interference from each MU-MIMO user from the ideal case. In comparison, the FDD system mostly depends on the separation of the main lobes between the users and the sidelobe levels of a chosen codebook beam pattern rather than the null placement. Since the gain from main lobes of a beam are less sensitive to phase errors~\cite{Madhow2012} the FDD system degrades less as a percentage from the peak. The results from simulations confirm this effect with the percentage loss in sum-throughput at \(20\degree\) RMS at \(20\%\) with TDD and \(3\%\) with FDD.  MF precoder performance described later in Fig. \ref{fig:thr:mf} with TDD also shows degradation similar to FDD case due to similar reasons. 

Fig.~\ref{fig:error} depicts the performance of a FD-MIMO system in the presence of magnitude or amplitude errors. The assumptions for simulations are similar to the phase error case. The system is assumed to have perfect phase alignment and magnitude error is assumed to vary from 0 to 9 dB where 0 dB corresponds to zero magnitude error in linear scale. The magnitude error is modeled as an additive zero-mean Gaussian random variable with a standard deviation of~$10^{{\sigma^{2}_{\dB}}/{10}} - 1$. Fig.~\ref{fig:error} shows that when the magnitude error increases, the throughput gain from using more antenna ports decreases significantly. More specifically, Fig.~\ref{fig:mag:fdd:1} and \ref{fig:mag:tdd:1} show that the advantage of using more digital ports disappear when the error goes beyond 4 dB RMS. Sum throughput of 32 ports system with RMS magnitude error greater than 1.5 dB performs worse than a perfect 16 port system. Similar to phase error case scaling to higher number of ports only improves performance if the amplitude errors do not degrade with scaling. Comparing FDD and TDD systems the TDD system is more sensitive to magnitude errors. Its performance degrades rapidly from the ideal scenario and comes closer to FDD at errors greater than 1 dB RMS.  

Since hardware impairments typically cause a combination of phase and magnitude errors, the individual limits for acceptable performance need to be more conservative. Hence, limiting phase error to  \( 10 \degree \) RMS and magnitude error \( 0.5 \) {dB} ensures the degradation is within \(10-15\% \). 

In Fig.~\ref{fig:thr:mf}, we compare the sum-throughput performance of ZF and MF digital precoders in the presence of phase and magnitude error with 32 ports and 4 users. The magnitude error is fixed at 1 dB and phase error is varied from 0\degree to 180\degree. With no phase errors, the ZF precoder outperforms the MF precoder by a large margin. The MF precoder only maximizes the energy to the target user, while the ZF precoder tries to minimize the interference from co-scheduled MU-MIMO users. As phase errors increase, performance of ZF precoder degrades rapidly and becomes comparable to MF. Therefore, it may be better to choose a MF based precoding that has much lower computational complexity if hardware impairments cannot be easily overcome with larger array sizes.

Fig.~\ref{fig:single:multi} compares the throughput of the single cell and multi-cell scenarios for both FDD and TDD systems versus phase and magnitude error with 32 ports and 4 users. Fig.~\ref{fig:single:multi} shows that the rate of degradation due to phase error in the single cell case is steeper than multi-cell one. Hence, the degradation from RF impairments has a reduced effect on performance degradation when interference from neighboring cells is also considered.

\begin{figure}[t!]
\centering
\includegraphics[height=0.48\textwidth, angle=-90]{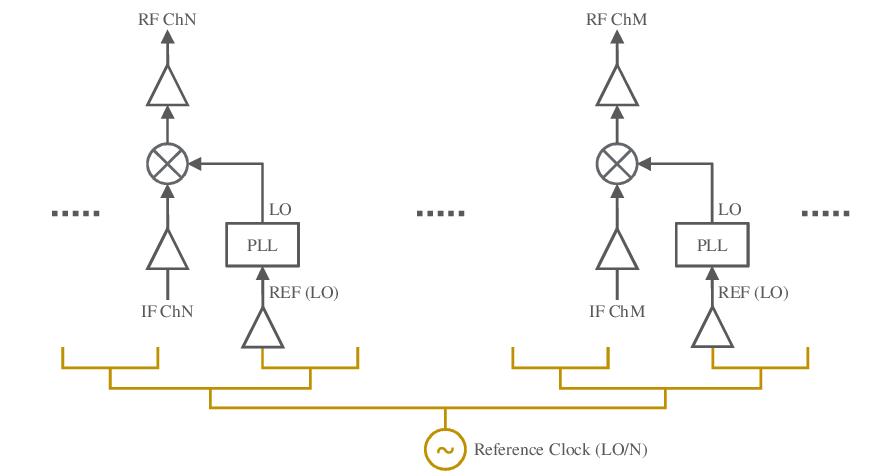}
\caption{PLL-based LO distribution}
\label{fig:lo:pll} 
\end{figure}

\section{Characterizing Phase and Magnitude Stability} \label{PhMagVarChr}
RF chains in an active antenna array can have different frequency responses that also vary with time during operation. Response that differs statically are typically affected by factors such as manufacturing tolerances and mutual coupling between antenna elements which can be measured and compensated using test equipment along with near/far field measurements in a factory setting before deployment~\cite{Willwerth1995,Kuehnke2001}. On the other hand, the dynamic variations require correction during operation to avoid performance degradation. There is a lack of experimental characterization of phase and amplitude variation in large active antenna arrays designed for cellular deployment which can help understand the practical limits of performance or the design tradeoffs. Although past research describes relative calibration methods there is not much experimental data confirming their efficacy in real world environment with continuous transmission. Hence in this section we investigate the major sources of phase and amplitude variations that include LO phase drift and variation in frequency response across RF chains due to internal temperature variation. We provide measurements from both lab bench tests and from a fielded system to characterize the behavior. 

\subsection{Phase Variation due to LO Phase Drift} \label{PhVarLo}

Change in phase response due to unequal phase variation at LO output used for mixing at each RF chain is one of the major sources of phase error in active antenna arrays. Although multiple LO architectures have been proposed in prior massive MIMO research we show that their applicability still remains challenging for practical deployments.

Three different approaches for LO generation and distribution have been suggested in the literature for massive MIMO: i) separate LO (SLO) where a LO signal is generated by an independent oscillator locally for each transceiver~\cite{Debbah2015, Krishnan2016}; ii) phase locked loop (PLL) based where a low frequency reference is first distributed to each transceiver and a PLL is then used to generate desired LO frequency~\cite{PuglielliBwrcPnBF2016}; and iii) common LO (CLO) where the desired high frequency LO signal is directly distributed from a common source to all transceivers~\cite{Krishnan2016}.  

The SLO architecture was shown to be feasible for practical deployment and to outperform CLO in certain cases~\cite{Debbah2015,Krishnan2016}. However such a result is likely due to the assumptions made including: (i) very short time duration between uplink training and downlink data transmission of about 100-10,000 symbols in the range of 0.25 ms; (ii) only considering the effect of the time averaged power of LO phase error or phase noise ignoring the effect of LO phase drift over time~\cite{Galton2019}. In LTE and 5G new radio (NR), the elapsed time between uplink SRS based channel estimation or CSI reporting and downlink data allocation to a UE can extend up to many multiples of 5 ms period. Hence, practical implementations require LO stability over larger duration. 

Phase variation caused by an LO is generally modeled by a Wiener process where the variance of the phase error at time duration equal to the time elapsed between uplink channel estimation and data transmission is given by~\cite{FetweissTcom2007}
\begin{equation} \label{eq:phNoise}
    {\sigma_\phi^{2}(t,\tau) = 4\pi^2f_{c}^{2}c\tau = 4\pi^2f_{c}^{2}cT_{s} \frac{\tau}{T_{s}} = \sigma^{2}_{\phi {T_s}}l }
\end{equation}
where \(c\) is the phase noise constant of the free running oscillator~\cite{FetweissTcom2007}, \(\tau\) is the time elapsed between training and downlink transmission, \(T_{s}\) is the OFDM sample time duration, \(l= \frac{\tau}{T_{s}} \) is the number of samples occupying the elapsed time and \( \sigma^{2}_{\phi{T_S}}\) is the variance computed from the frequency domain phase error spectra of the LO. Analysis based on SLO architecture uses this method to model the LO phase error. This model based only on frequency domain phase error spectra is not very useful in capturing the short term time domain frequency stability of an LO~\cite{Galton2019} that results in phase drift over time especially with low cost crystal oscillators. VCXOs and TCXOs are commonly used in cellular radios~\cite{Rakon2009}. The effect of short term time stability of LOs on distributed-MIMO beamforming application was examined in~\cite{Kumar2014} and also experimentally verified  in~\cite{Brown2017}. These results confirm that independent LOs (TCXOs) drift out of phase in much less than 10 ms degrading beamforming gain and requiring frequent training that drastically increases overhead.

This discrepancy between the two models of LO phase error is further examined for cellular application. Consider an LTE air-interface with a bandwidth of 20 MHz and a carrier frequency of 2 GHz. In order to satisfy the EVM requirements for RF conformance~\cite{3GPP2017104} at individual transceiver array boundary (TAB) connector~\cite{3gppAAs2015} the contribution to EVM from LO is attributed to the frequency domain phase noise which is twice the one-sided phase error spectra of the LO. This region is dominated by white phase modulation noise~\cite{Rakon2009}. The phase error variance can be expressed as 
\begin{equation}
{\sigma^{2}_{\phi{T_S}} = 2 \int_{f_L}^{f_H} \mathcal{L}\big( f\big) df}    
\end{equation} 
where \( \mathcal{L}(f) \) is the one-sided power spectrum of phase error, \(f_L\) is 15 KHz and \(f_H\) is 10 MHz. In the case of LTE, the integrated phase error is typically limited to -40 dBc corresponding to an EVM of \(1\%\)~\cite{Minihold2013}. This is equivalent to a frequency domain integrated phase noise of \(\sigma^{2}_{\phi{T_S}} = 10^{-4}\ \mathrm{rad}^{2}\). The variance in phase error over an elapsed time of \(\tau = 1 {ms}\) from \eqref{eq:phNoise} for LTE air interface can be determined as $ {\sigma_\phi^{2}(t,\tau) = 10^{-4} \times 30720 = 3 \ \mathrm{rad}^{2}}$. This corresponds to a RMS phase error of \(99\degree\). In order to satisfy a RMS phase error requirement of \( 10\degree\) would require the integrated phase noise to be less than -60 dBc  using the frequency domain approach.   

Now if we consider the time domain short term stability of a TCXO oscillator the phase error variance of an LO output is typically characterized by Allan variance. It is the two sample variance of an oscillator output for a given gating period \(\tau\)~\cite{Galton2019}.  The state space model described in~\cite{Zucca2005} and used by~\cite{Kumar2014,Brown2017} uses a multiple Wiener processes with different variances to model phase error of a LO. Each Wiener process is mapped to an underlying dominant noise source that depends on the gating period considered for Allan variance measurement. Based on this for a time duration greater than \(100\mu{s}\) and less than a \(1s\) for a TCXO, the underlying noise is dominated by flicker phase and white frequency modulation noise~\cite{Rakon2009}. In this period the mean squared phase error for the Wiener process can be calculated as~\cite{Zucca2005}

\begin{equation}
    {\sigma_\phi^{2}(t,\tau) = 4\pi^{2}f_c^{2}\sigma_{Adev}^{2} (\tau) \tau^2}.
\end{equation}
For TCXOs the Allan deviation \(\sigma_{Adev} (\tau) \) is typically specified as \(10^{-9}\)~\cite{Vig1992} at a gating period of \(1{s}\). Between \(100 \mu s\) and \(1s\) the product \(\sigma_{Adev}(\tau) \tau\) is a constant~\cite{Galton2019}. Hence for an elapsed time of \( 1 {ms} \) the variance in phase error based on Allan deviation is given by
\begin{equation} \label{varPhAllan}
 {\sigma_\phi^{2}(t,\tau) = 4\pi^{2}f_c^{2}(10^{-9})^{2} = 158 \ \mathrm{rad}^{2}}.
\end{equation}
 This corresponds to a RMS phase error of \(720 \degree\). It is much higher than the value calculated using frequency domain approach. In order to satisfy the RMS phase error of less than \(10\degree\) the Allan deviation of the LO based on \eqref{varPhAllan} needs to be around \(10^{-11}\) which is only satisfied by higher quality LOs like OCXOs~\cite{Tufvesson2014} which make the system very expensive for SLO type implementation. 

Another problem with SLO architecture relates to degradation due to carrier frequency offset correction. Solving this requires either a complex receiver based on channel estimation~\cite{Debbah2015, Mehrpouyan2012} or multiple frequency tracking loops at the UE~\cite{PuglielliBwrcPnBF2016}. In order to avoid this the PLL based LO generation method is proposed which is shown in Fig.~\ref{fig:lo:pll}. This solution makes the close-in phase noise, which is dominated by the low frequency reference distributed to each PLL and responsible for short term instability, correlated across all LOs. This makes the relative phase drift between RF chains more stable than the SLO case. This also simplifies the carrier frequency offset correction at the UE by only requiring a global frequency tracking loop. The PLL based method makes the system frequency coherent but not fully phase coherent. Hence for longer time duration's greater than a few $10$s of minutes the phase deviation between two RF chains can degrade as much as \(20\degree\)~\cite{Keysight2014}. This is good enough for LTE and NR systems where training and CSI feedback for data channels are carried out within $100$ ms. In the case of broadcast channels and beamformed CSI-RS transmission that operate without knowledge of CSI and require specific radiation patterns the main lobes are less sensitive to phase and amplitude error. Hence a PLL based LO solution that restricts the phase error to $10 \degree$ RMS at any time should be sufficient.

In practice however, cabling or transmission line used for distributing the reference signal across large electrical distances of the antenna array is more challenging since it is highly susceptible to attenuation and propagation delays. Existing solutions either use distribution PLL ICs or clocks~\cite{ADIRefDist2018,Tufvesson2014} to distribute the reference signal or use digital high speed distribution methods with intricate real-time triggering and clock buffer management mechanisms~\cite{TIJesd2015, ADIJesd204B}. However these solutions have multiple design challenges in achieving phase coherency including, meeting EVM limits due to additional noise from intermediate PLL ICs, frequent re-synchronization and re-timing for time alignment due to high susceptibility to temperature variation, maintaining symmetry between digital traces and extensive software development required for trigger and clock management. Hence PLL based solutions are challenging to implement for for practical cost effective FD-MIMO systems.

The remaining CLO architecture which is the only method known to achieve phase coherency better than $1 \degree$ RMS performance in a lab setup~\cite{Hall2014,RhodeSchwarz2016,Keysight2014} is very difficult to implement across a large array. Distributing a high frequency signal (\(>1 \mathrm{GHz}\)) is more challenging than a low frequency signal discussed in PLL method due to higher sensitivity to impairments at smaller wavelength~\cite{ADIClkSkew2019}. 

Hence there is still a need for a better LO distribution method that provides adequate coherency with low re-calibration overhead. In the next section we present a novel LO distribution method called Bi-directional Signaling (BDS) that performs similar to CLO approach with very low overhead. 

\subsection{Phase and Amplitude Variations in the Frequency Response of RF Chains} \label{PhAmpFrespRf}
\begin{figure}[t!]
\centering
\begin{minipage}[b]{0.45\textwidth}
\centering
\includegraphics[width=\textwidth]{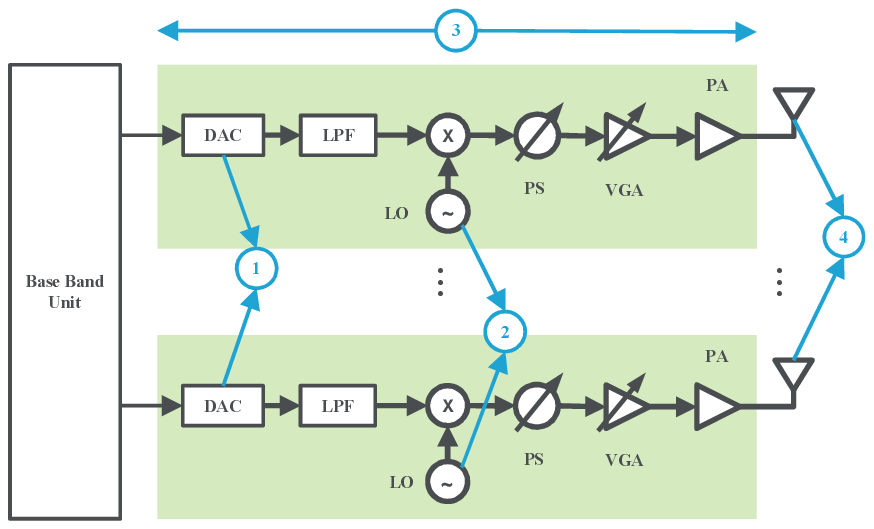}
\caption{Sources of calibration errors in an active antenna array on the transmit side.}
\label{fig:source}
\end{minipage}\hfill
\begin{minipage}[b]{0.5\textwidth}
\includegraphics[height=\textwidth, angle=-90]{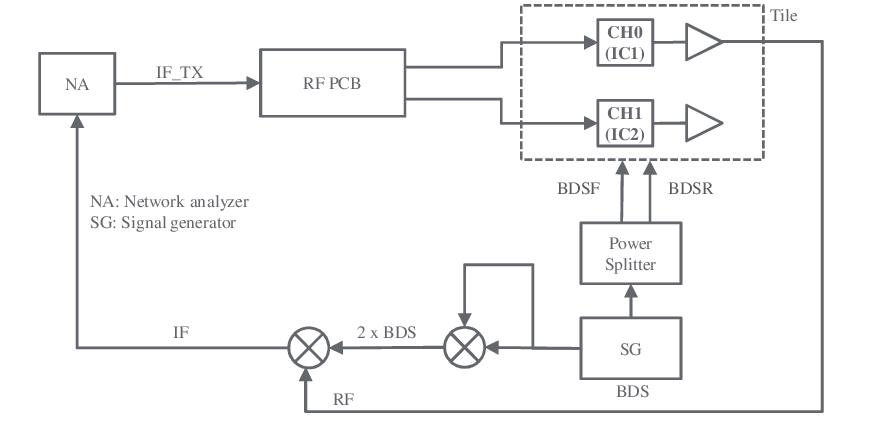}
\caption{Lab test Bench setup to measure phase and magnitude error.}
\label{fig:bench} 
\end{minipage}
\end{figure}

Fig.~\ref{fig:source} shows simplified typical transmitter chains of an active antenna array. Receiver chains are similar and are not shown here for simplicity. Notice that despite the usual label of “digital radio” blocks, these transmitter chains are made of pure analog/RF circuits such as amplifiers, mixers, filters, electronic attenuators, voltage regulators, data converters, etc.  Fabrication process, bias and temperature variations affect the frequency responses of most analog/RF components; therefore, phase and amplitude errors will normally occur in any of these analog/RF blocks. These errors are in addition to the LO phase drifts we already covered. 

There are many techniques for minimizing or even eliminating the phase magnitude errors occurring in analog/RF components but usually these techniques come with a heavy penalty in cost. Process variations can be compensated with offline factory calibrations. Internal temperature and voltage variations affect group delay, phase distortion and gain of amplifiers; group delay of RF filters; sampling clocks of data converters introducing sampling error and frequency response across cabling, connectors and couplers used for LO and IF/RF distribution~\cite{RhodeSchwarz2016}. Voltage fluctuations can be limited by using higher grade power supplies. However internal temperature variation is highly dependent on thermal design which is very challenging for macro-cell active antenna systems with sealed enclosures. Hence it is important to characterize the phase and amplitude response to temperature variations across the array during operation under real-world environments. This is important to understand calibration requirements and assess realistic gains from massive MIMO.

Prior work have characterized the effect of temperature fluctuation on smaller phased arrays with 8 RF chains under controlled lab environment~\cite{Beach1998,Brauner2003,Aghvami2007}. A commercial FD-MIMO active antenna array requires much higher number of RF chains (16-128) generating heat. The active components are also required to operate in a sealed enclosure with limited ventilation devoid of fans ~\cite{Honcharenko2019}. To the best of the authors knowledge, there have been no prior characterization of temperature variations on any practical large antenna arrays system designed for cellular application or FD-MIMO. Therefore in this section we provide measured data from an active antenna array implementation called High Definition Active Antenna System or HDAAS to help characterize the effect of internal temperature on frequency response.

\begin{figure}[t!]
\centering
\subfloat{\label{fig:cal:temp:phase}\includegraphics[width=0.48\textwidth]{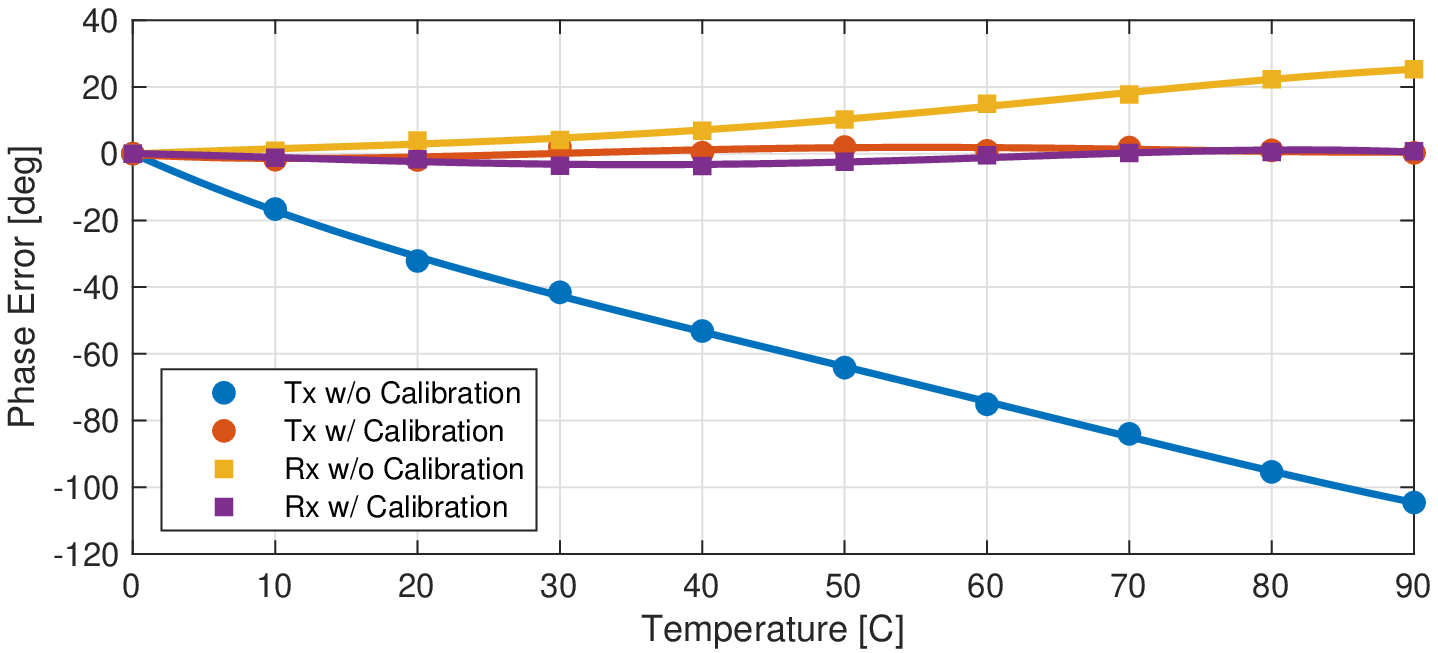}} 
\subfloat{\label{fig:cal:temp:mag}\includegraphics[width=0.48\textwidth]{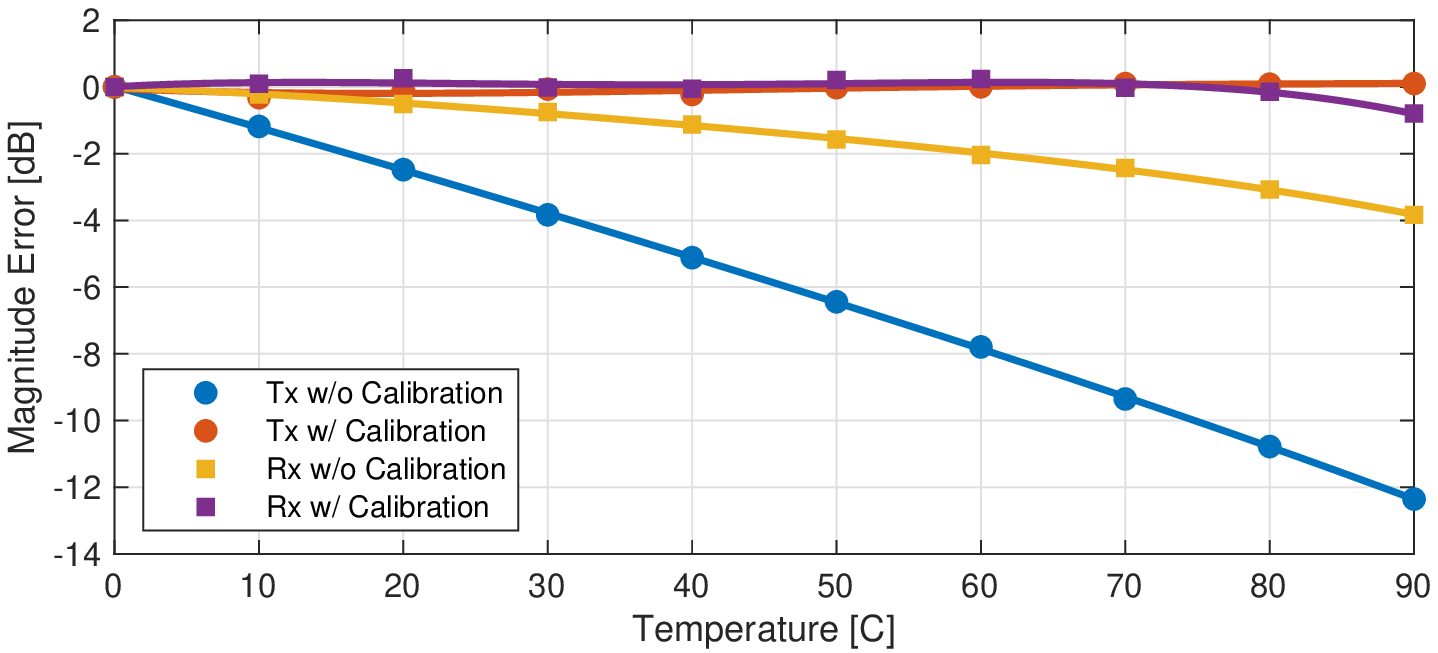}} 
\caption{Variation in phase and magnitude response with temperature}
\label{fig:cal:temp}
\end{figure}

\begin{figure}[t!]
\centering
\subfloat[Absolute temperature]{\label{fig:temp:time:absolute}\includegraphics[width=0.48\textwidth]{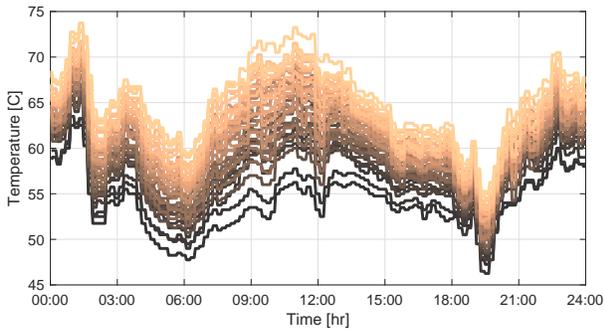}} \hfill
\subfloat[Relative temperature]{\label{fig:temp:time:relative}\includegraphics[width=0.48\textwidth]{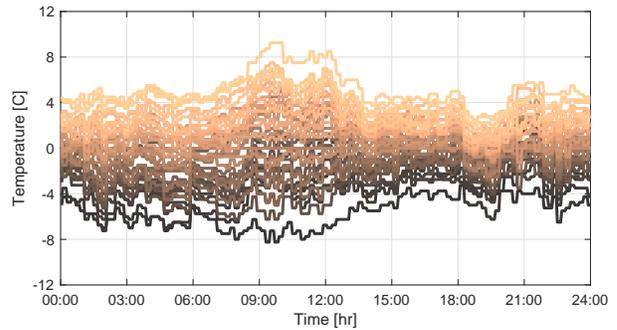}} 
\caption{Temperature recorded across 48 RF chains over time.}
\label{fig:temp:time}
\end{figure}

\begin{figure}[t!]
\centering
\subfloat[Absolute temperature]{\label{fig:temp:phase:absolute}\includegraphics[width=0.48\textwidth]{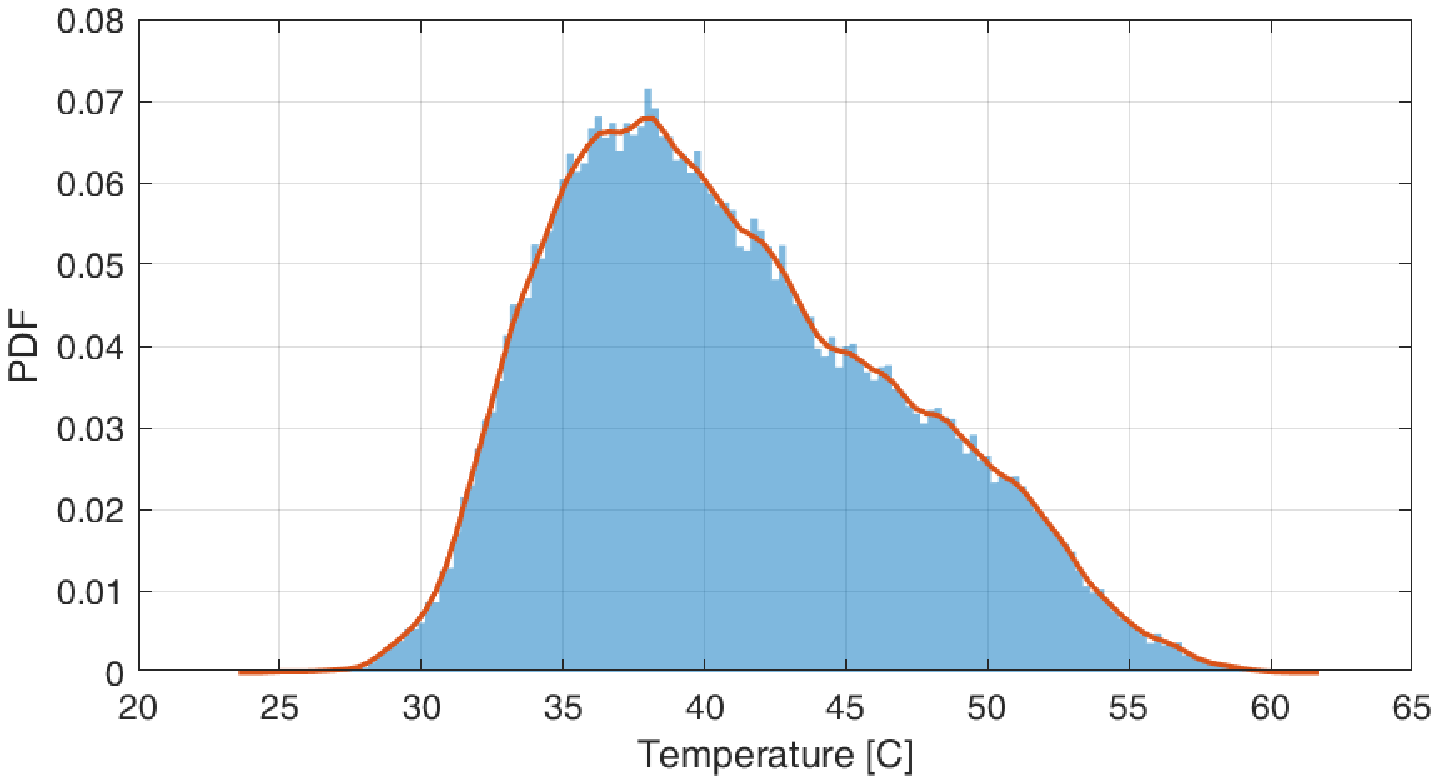}} \hfill
\subfloat[Relative temperature]{\label{fig:temp:phase:relative}\includegraphics[width=0.48\textwidth]{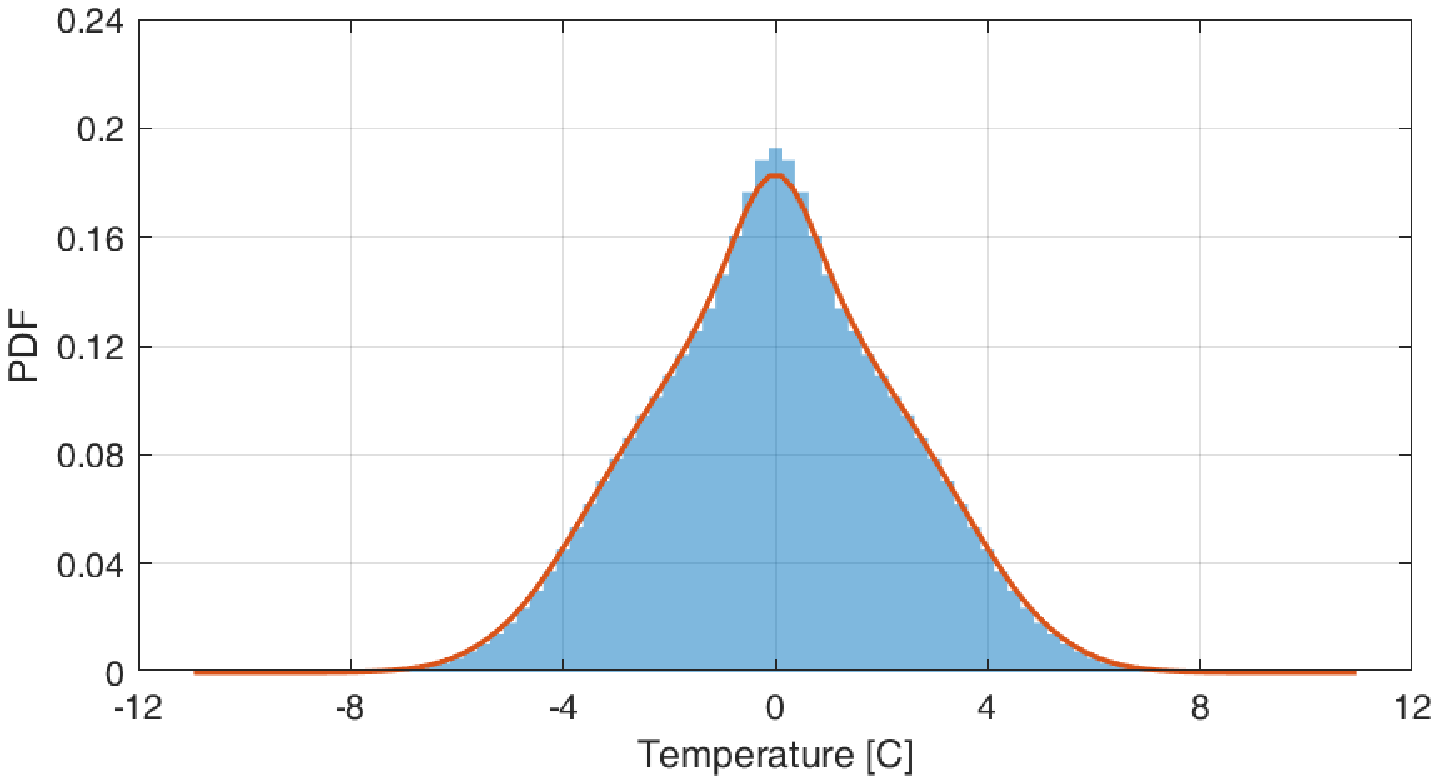}} 
\caption{Distribution of absolute and relative temperature.}
\label{fig:temp:phase}
\end{figure}

The block diagram describing the architecture of HDAAS is shown Fig.~\ref{fig:arch}. More details about the design are provided in the next section. The modular section that contains all the distributed RF front end electronics excluding the data converter, called a 'tile', resides behind every antenna element. Each tile supports RF chain for two cross-pol elements.  A temperature sensor in each tile records the temperature used for analysis. We first measure the variation in output phase and power at the output of a tile at different temperatures using thermal pads under a tile in a lab set up shown in Fig.~\ref{fig:bench}. The results are plotted in Fig.~\ref{fig:cal:temp}. In the case of transmit path the power output decreases almost linearly with temperature reducing by \(2 \mathrm{dB}\) for every \(20 \degree \mathrm{C}\) increase in temperature. The phase changes by \(20 \degree\) with \(20 \degree \mathrm{C}\). In the case of receive path the variation in received power gain and phase is much lower compared to transmit chain. The gain reduces by about \(2 \mathrm{dB} \) with an increase of \(60 \degree \mathrm{C}\) and the phase increases by \(29 \degree\) with an increase of \(90 \degree \mathrm{C}\). 

Next we plot the temperatures recorded across 48 tiles during operation over one day period from a field deployment of the HDAAS on a tower top in a LTE macrocell network carrying commercial traffic Fig.~\ref{fig:temp:time:absolute}. Each line in the plot corresponds to the temperature related to a tile corresponding to an antenna element. A temperature recording is made every 3 minutes. The tile temperatures across the array appear to move in the same direction as the average across all the tiles. The relative temperature variation with respect to a reference antenna is shown in Fig.\ref{fig:temp:time:relative}. The relative temperature on any tile with respect to a specific reference tile is as high as \( \pm 9 \degree \mathrm{C}\). Hence depending on the reference tile chosen the difference can be as high as \(11 \degree \mathrm{C}\) shown in Fig. \ref{fig:temp:phase:relative} which is the probability density function (PDF) of the relative temperature between any two tiles measured over a long duration of 2 weeks. The internal temperature gradient can depend on several factors including size of the array, thermal design, components used including the number of digital processors used, ambient temperature outside the array and in some cases whether sunlight hits part of the array. The PDF of absolute temperature variation on a single tile for a period of 1 month is shown in Fig. \ref{fig:temp:phase:absolute}. The internal temperature can be seen to vary as much as \(30 \degree \mathrm{C}\). This data shows the importance of characterizing non-ideal behavior of large arrays in practice to understand what is realistically possible in terms of compensation and capacity improvement even though the data is specific to HDAAS. 

Calibration is required to compensate for deviation from ideal behavior. The frequency at which calibration routines need to be activated as seen from the temperature measurements above depends on the rate of relative internal temperature fluctuation in Fig.~\ref{fig:temp:time:relative}. Based on results from bench tests in Fig.~\ref{fig:cal:temp} relative deviation after initial calibration can use \(3 \degree \mathrm{C}\) as a threshold for activation. This limits maximum phase error within $6 \degree$ and magnitude error to $0.6 \mathrm{dB}$. On the other hand if a calibration on each RF chain is carried out every 30 minutes it should satisfy the above threshold requirements based on Fig. \ref{fig:temp:time:relative}. However, it should be noted that these recommendations on calibration frequency are specific to HDAAS array construction. The internal temperature variations can be much worse in other types of array designs making it harder to calibrate and achieve ideal performance. Based on the two major sources of multiplicative noise and their behavior discussed in this section we next discuss solutions that solve coherent LO generation and calibration.

\section{Methods for LO Generation and Absolute Calibration} \label{calibration}
In this section, different methods to generate LOs are described. A good overview of various methods proposed in literature for calibration of massive MIMO systems is provided in~\cite{Tufvesson2017} and references thereof. Calibration methods developed so far are only applicable to reciprocity based TDD system. These cannot be applied to FDD FD-MIMO and are mostly applicable to element level digitized array or sub-array hybrid beamforming. 

In typical LTE macrocell deployment the remote radio head (RRH) implements the RF front end on a tower top. The digital baseband unit (BBU) is either situated at the bottom of the tower or in a data center connected to RRH over fiber. The RRH and BBU are required to operate coherently. This is achieved by making the RRH time synchronized with the BBU using a synchronization protocol over common public radio interface (CPRI). Hence any calibration method implemented needs to maintain coherency end-to-end from the antenna to the baseband unit (BBU)~\cite{Ericsson2016}.

In the case of relative calibration~\cite{Zhong2012,SamsungJsac2017FDMimoProto}, calibration coefficients are calculated as the ratio of the frequency responses measured between a reference RF chain and other RF chains can be calculated as \(c_i = \frac{t_{i} r_{ref}}{r_{i}t_{ref}} \) where \(t_i\) and \(r_i\) are the transmit and receive frequency responses of \(i^{th}\) RF chain and \(t/r_{ref}\) is the reference Rf chain. if \(h_{k \mapsto i}\) is the channel response between the \(k^{th}\) user and \(i^{th}\) receive antenna. Then the CSI estimate using the calibration coefficient is computed as

\begin{equation} \label{eqCSITi}
\hat{h}_{i \mapsto k} = c_{i}r_{i}h_{k \mapsto i} = \frac{t_{i} r_{ref}}{r_{i}t_{ref}}r_{i}h_{k \mapsto i} =  \frac{r_{ref}}{t_{ref}}t_{i}h_{k \mapsto i}.
\end{equation}
It can be observed that \(t_{i}h_{k \mapsto i}\) is the desired CSI and the \(\frac{r_{ref}}{t_{ref}}\) is a constant offset that is applied to all RF chains. Hence using the calibration coefficient  the modified CSI estimate only differs from the desired CSI by a constant offset which is compensated in downlink by channel estimation at the UE using demodulation reference signals (DMRS). The relative phase and amplitude between transmit chains required to determine the linear precoder is preserved by relative calibration. The performance of the system heavily depends on the accuracy of the calibration coefficient. The system performance is sensitive to variation in temperature that modifies frequency response of RF chains or inaccuracy in measurement due to errors introduced by calibration network.

In the case where channel feedback from UE is used to determine the precoder for FD-MIMO, the relative phase and magnitude between the transmit chains need to correspond to the chosen precoder. Calibration needs to also make sure that the relative phase between antennas during channel feedback is maintained until data transmission. The beamforming gain and user separation in this case depends on the array aperture/size and coherence achieved. This requires the array to maintain a stable array phase center relative to which the phase deviations can be compensated against. This array phase center also needs to be phase coherent with the clock used in baseband processing to meet end-to-end link coherency. Relative calibration cannot be used in such scenarios. Hence another approach is to perform absolute calibration. First a good phase and magnitude reference needs to be established that is invariant over temperature and time. Before the array is set into operation mode, the relative phase and magnitude response of each RF chain is determined with respect to this system reference and a correction factor is applied to achieve desired performance under stable conditions. The common system reference acts as the array phase center. In the operation mode periodic or on-demand measurements are performed to identify deviations from the initial calibration values which is then compensated as phase offset or gain correction to the respective chain. This makes the transmit and receive arrays revert back to pre-determined offsets and only additional precoder related phase and magnitude offsets need to be further applied. This method can be applied to both channel feedback as well as channel reciprocity based FD-MIMO systems. The main challenges of this approach include creating a stable reference for array phase center and creating a stable calibration network to perform consistent measurement of the deviations in frequency response.

Here we provide some of the requirements that need to satisfied for calibration in a cellular deployment and discuss the drawbacks of some of the methods described in literature. Meeting following requirements are highly desirable when sounding is used for calibration: i)  Sounding cannot involve feedback from the UEs as the feedback overhead will overwhelm the transmission system. Hence self-calibration of the array is preferred. ii) Due to the compact space available~\cite{Honcharenko2019}, over-the-air calibration based on sounding with an external reference antenna is not feasible. iii) In an FD-MIMO system calibration should ideally be run simultaneously with transmission without causing any degradation to transmitted or received signals of the air-interface. This means meeting all RF conformance test requirements with respect to transmitter and receiver RF conformance~\cite{3GPP2017104} that include ACLR, EVM and receiver sensitivity limits for individual RF chains and limits for beamformed over the air measurements etc. Any interference from sounding affects spectral efficiency since time-frequency resources will need to be blanked out from active transmission during sounding. This also results in additional overhead  for tight coordination between MAC layer resource scheduler and calibration process. Since internal temperature variations can be unpredictable it can be difficult to assure deterministic performance.

Next we examine the applicability of commonly suggested methods in massive MIMO literature based on above requirements:
i) Channel sounding based on pilot signals or a tone  injected to determine calibration coefficients for relative calibration~\cite{SamsungJsac2017FDMimoProto} requires the sounding signal to be distributed over a symmetrical network. This means sounding signal power is split equally to ensure similar signal distribution to each RF chain and maintaining identical response as observed during transmission. A circuit based calibration to create a symmetric network requires routing over transmission lines, directional couplers, switches and dividers with careful considerations to avoid interference or superposition which is not easy. Implementing the calibration network under tight space constraints with signal going across multiple boards of digital and analog front end components~\cite{Honcharenko2019} can be complex. High isolation required between calibration network and receiver during data reception can be very hard especially for compact arrays which can necessitate blanking active transmission time-frequency resources during calibration. In the case of absolute calibration method the symmetric network requirement can be relaxed since offline calibration can be used to quantify and compensate the mismatches.  

ii) Channel sounding based on mutual coupling based over-the-air measurement between adjacent antennas by re-using existing RF chains~\cite{KaltenbergerFmwkCal2018}. There is a lack of experimental validation of the efficacy of these methods in compact enclosures~\cite{Honcharenko2019}. Relative calibration using mutual coupling again requires sequential channel sounding over different pairs of transmit and receive chains in order to avoid mutual interference. This method can also have similar drawback as circuit based calibration due to poor isolation that results in higher interference in uplink. In addition signal transmitted for channel sounding can saturate a receiver due to coupling and also pose challenges satisfying downlink RF conformance requirements related to ACLR and EVM.

It is clear that calibrating without affecting active transmission is highly desirable and difficult in practice. Hence we next describe a novel calibration method that takes advantage of the availability of an array phase center derived from the phase reference of a LO distribution method described earlier. In order to describe this method, we first describe the architecture of HDAAS system, which is a full connected hybrid beamforming system tested and fielded in a commercial FDD LTE network.

\section{A New Approach for Coherent LO Generation and Array Calibration} \label{calibHdaas}
In this section we describe a new approach to manufacturing a super low-cost coherent Massive MIMO system based on our large active array concept called High Definition Active Antenna System or HDAAS. The key system properties of HDAAS are a) large RF coherent active aperture with no data-flow interruptions for calibration, b) multiple digitization of the full aperture with programmable independent weights for each antenna element and each digitization instance, c) agile beamforming and switching operations and d) flexible control points including independent operation, BBU control and cloud-based control.  HDAAS has the least amount of HW of any Massive MIMO system for optimum cost and power dissipation because HDAAS captures the degrees of freedom of the channel with the least number of radio chains.

\begin{figure}[t!]
\centering
\includegraphics[height=\textwidth, angle=-90]{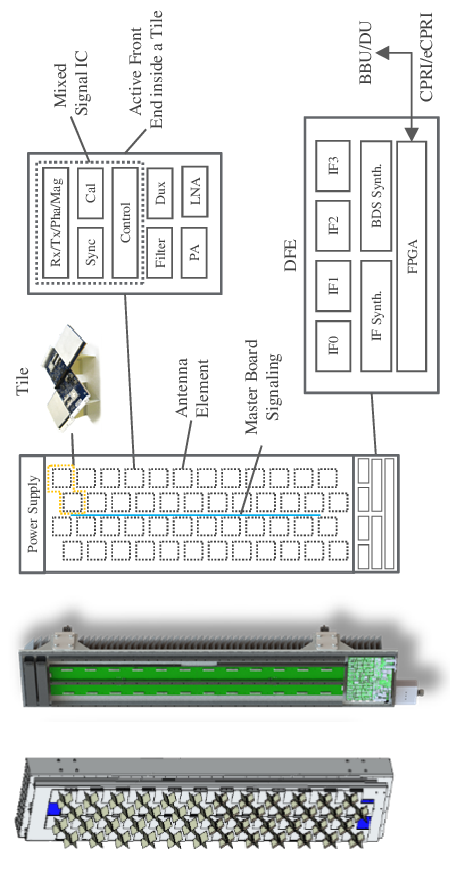}
\caption{HDAAS architecture}
\label{fig:arch}
\end{figure}

\subsubsection{HDAAS Architecture} \label{calibHdaasIfRf}
The high-level architecture of HDAAS is shown in Fig.~\ref{fig:arch}, constructed of power supplies and three main circuit boards: an active IF/RF/digital front-end board called a Tile, a large passive board called Master-Board (MB) and a Digital Front-End (DFE) board. The Tile contains standard radio circuits such as antenna elements, connectors, RF couplers, RF filters, up/down conversion mixers, phase rotators, variable-gain amplifiers, PAs, LNAs, power regulators, memory, digital microcontroller and novel mixed signal circuits for RF synchronization and system calibration. Most of these standard and novel circuits are integrated into a small, custom mixed-signal IC fabricated in an old, very low-cost Si technology (currently 180 nm BiCMOS). This integration is essential because HDAAS has a very large number of phase rotators and variable gain amplifiers supporting each transmit and receive beam in each Tile. Typically, a Tile contains circuits for two cross-polarized active antenna pairs, but other Tile designs supporting more active antenna pairs are possible. 

The MB contains transmission lines for analog IF signals, LO signals, a few synchronization/calibration signals, DC supply lines and digital control signals. It also contains connectors appropriately placed so numerous Tiles and one DFE can be plugged in to construct the HDAAS.

The DFE HW/FW functionality includes the digital link with the rest of the Massive MIMO system (CPRI to BBU in 4G or eCPRI to decentralized unit (DU) in 5G), the pre-data-converter digital processing (including radio unit (RU) digital functionality in 5G), digitization at IF for all channels, post-digitization analog signal conditioning, crest-factor reduction (CFR), and system synchronization/calibration support HW. The DFE SW functionality includes HDAAS control and management, beamforming, beam operations, and management of system calibration to maintain RF coherency over the entire HDAAS aperture at all times without any communication channel overhead. Typically, the DFE includes one FPGA with processors inside but other designs are also possible. 
 
When the HDAAS assembled as discussed above is powered up, the DFE has full and accurate digital control over the entire system including over the phase and magnitude settings of each radio signal path in each Tile. These settings can be changed every millisecond or even faster in full synchronicity with the channel data flow. The system RF calibration may be programmed to run continuously or as often as necessary in the background to maintain RF coherency over the entire aperture. The DFE control may run locally as an independent process,  may be provided to the FD-MIMO baseband unit (BBU/DU) or may be transferred to the element management system (EMS) that can reside in the cloud.  Combination of these three operation modes are also possible.

\subsection{System RF Synchronization}
One of the most important design features of HDAAS is the accurate RF synchronization (within a few degrees of error) of the Tiles with respect to each other. An RF system reference clock in the DFE is transmitted over the MB serially such that two instances of the clock arrive at each Tile over two separate paths. These paths are constructed such that the average arriving time between the two paths is a constant \cite{banu2012method}, even on a low-cost MB. This is accomplished by placing these paths on the MB very close to each other such that any local variations of the board materials, traces and temperature cancel each other to a high order. A simple version of this technique was used in the VLSI clock distribution method described in \cite{Banu2006VlsiClk}.
​

A special PLL circuit on the Tile extracts the average time of the two paths with high accuracy so each Tile generates a local clock signal, which by construction is exactly in phase (within a very small error) with the local clocks of all the other Tiles. Since these clock signals are already at RF, there is no need to use local frequency multiplication (often used in PLL clock generation), which is notoriously prone to generating large phase errors. Once the Tiles are thus synchronized at RF, the problem of maintaining RF coherency over the entire aperture (all Tiles) is simplified significantly. For example, all LO signals on the Tiles are derived from this local RF clock signals and thus are automatically in almost perfect phase alignment and with very low phase noise 

\begin{figure}[t!]
\centering
\begin{minipage}[b]{0.48\textwidth}
\includegraphics[width=\textwidth]{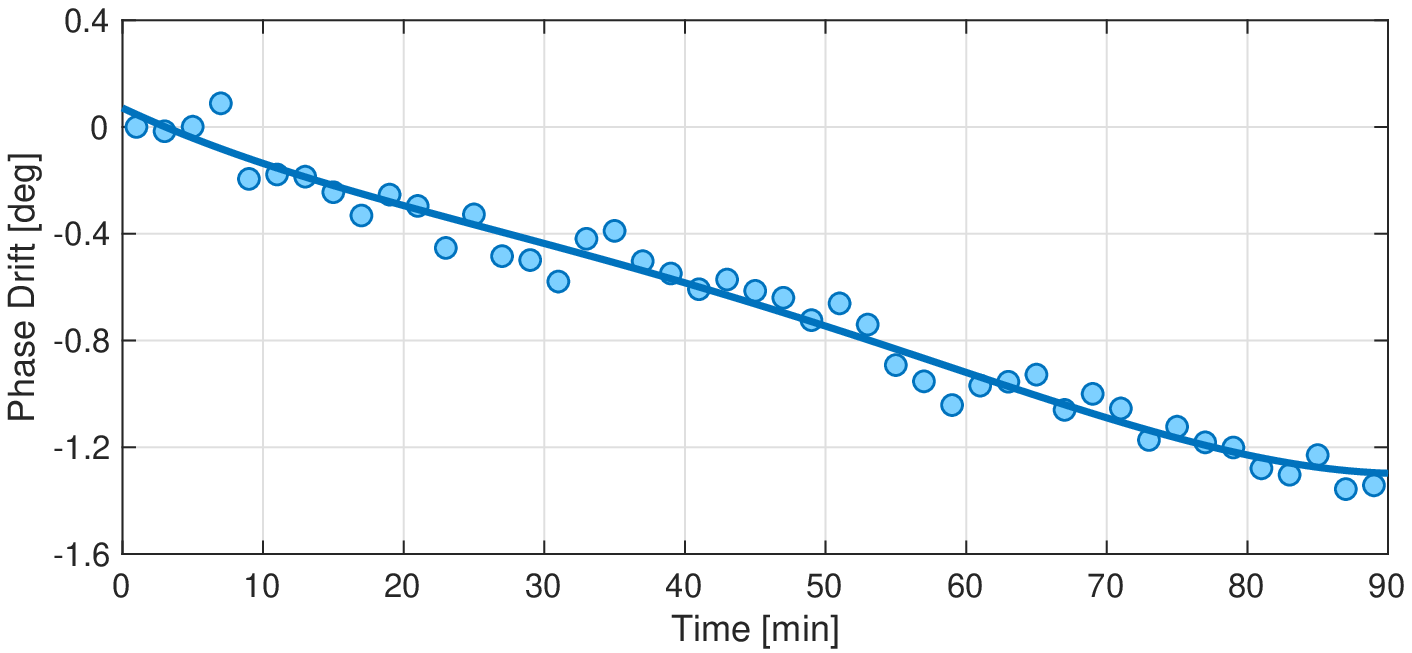}
\caption{PLL phase drift over time with BDS method.}
\label{fig:lo:time}
\end{minipage} \hfill
\begin{minipage}[b]{0.48\textwidth}
\includegraphics[width=\textwidth]{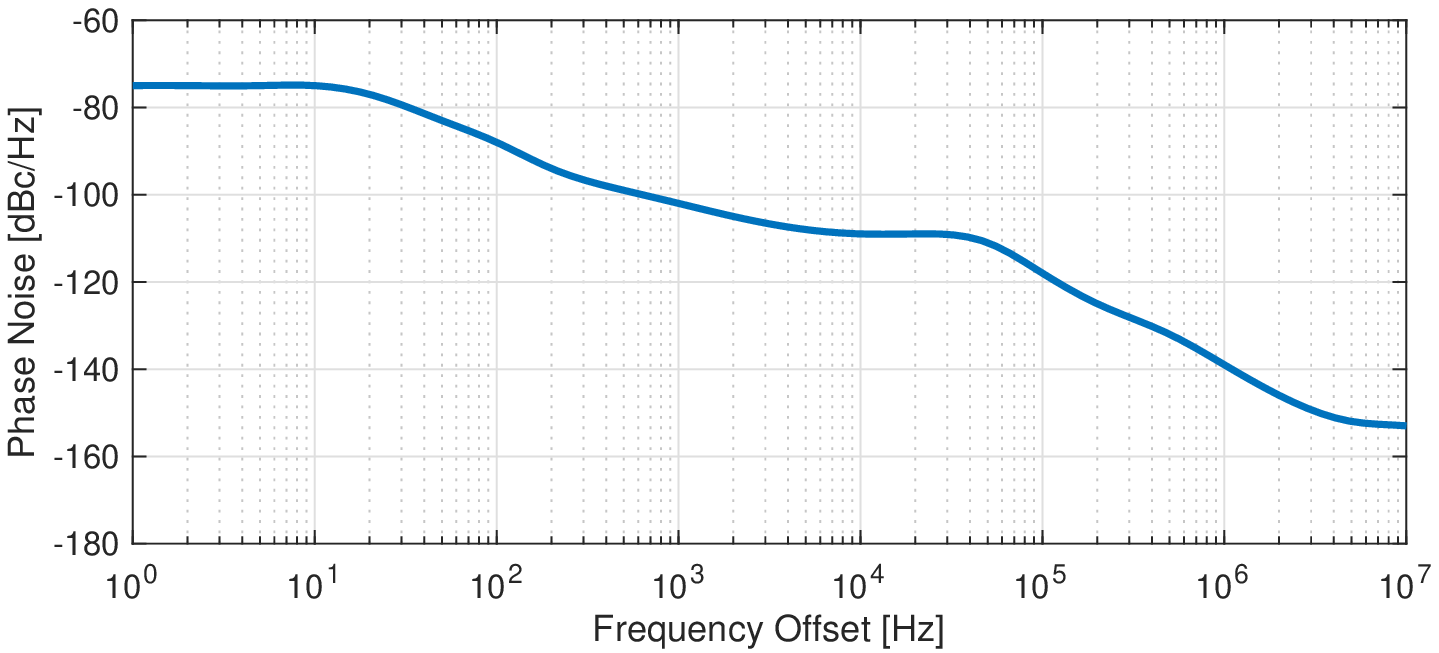}
\caption{Phase noise power spectral density with BDS method}
\label{fig:lo:phnoise} 
\end{minipage}
\end{figure}

Fig.~\ref{fig:lo:time} shows the measured LO phase variation of this method over a time period of 90 minutes at an ambient temperature of 21\degree C. The total phase drift due to secondary effects in the Tile is only 1.3\degree, which is insignificant and is removed during Tile calibrations (see below). Clearly, this method performs as well as the CLO method mentioned earlier. An additional significant advantage is a total  insensitivity to temperature variations, which means the LO distribution does not need re-calibration like the PLL or CLO approaches. The power spectral density of the LO signal extracted is shown in Fig.~\ref{fig:lo:phnoise}. The integrated phase noise is about −83 dBc which is well below −60 dBc discussed earlier.

Therefore, given the degree of phase coherency achieved with our  low-cost method for LO generation, it is an attractive solution for FD-MIMO array implementation. This method may be applied to all other array architectures. It is also possible to use this method for coherent clocking of  high speed data converters.

\subsection{Tile Calibration}
The Tiles contain extensive standard-quality, low-cost IF and RF circuits, which are subject to large fabrication and operational drifts due to temperature, humidity, aging, etc. The brute force solution, often adopted by classical military phased arrays, would be to replace these standard-quality low-cost circuits with very high-performance but very expensive components having stable electrical characteristics. Additionally, we would have to employ expensive high accuracy factory calibrations. Instead, HDAAS uses cell-phone quality low-cost components and coarse low-cost factory calibrations. The key to obtaining and maintaining RF coherency is a sophisticated background Tile calibration method, which monitors with high precision the RF path drifts in phase and magnitude and corrects them automatically. Each active antenna is electrically removed from the system one at a time, is calibrated for all RF paths and reconnected to the system. This system includes HW mechanisms in the Tile for phase and magnitude drift sensing,  firmware in the Tile for correcting the drifts and software in the DFE implementing the calibration management algorithm. Notice, that unlike the typical Massive MIMO systems developed in the industry, HDAAS can calibrate one or a few active signal paths at a time without the need to interrupt the operation of the rest of the system. When one or even a few Tiles are removed electrically, the system performance degradation is practically imperceptible.  This is also the consequence of the fact that the differential electrical characteristics of the MB traces are quite insensitive to operational drifts. 

\subsection{MB Calibration}
As the MB is fabricated in a low-cost PCB technology, the various transmission lines it contains are not guaranteed to be mutually matched after fabrication. For proper HDAAS operation it is necessary to make sure all IF paths in the MB are equal in electrical length. This is achieved with a similar calibration scheme to the Tile calibration. The only difference is that while the Tiles must be calibrated very often (e.g., every half hour), the MB calibration is typically done only once at system boot time. As mentioned earlier, the MB differential drifts with temperature and other operational conditions are insignificant. We have operated HDAAS units over long periods of time including very hot summers and very cold winters without re-calibrating the MB with no  measurable performance degradation.

\subsection{Lab and Field Tests}

\begin{figure*}[t!]
\centering
\subfloat[Near-field]{\label{fig:beam:a}\includegraphics[width=.48\textwidth]{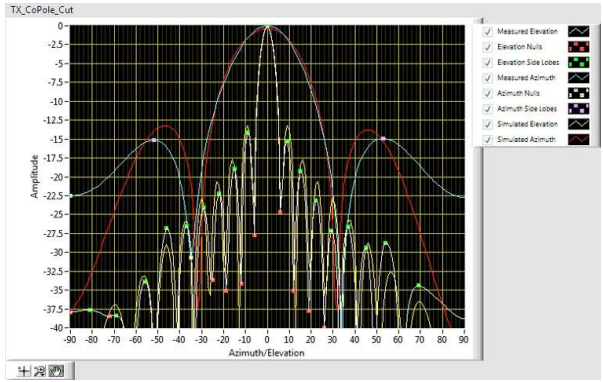}} \hfill
\subfloat[Far-field]{\label{fig:beam:b}\includegraphics[width=.48\textwidth]{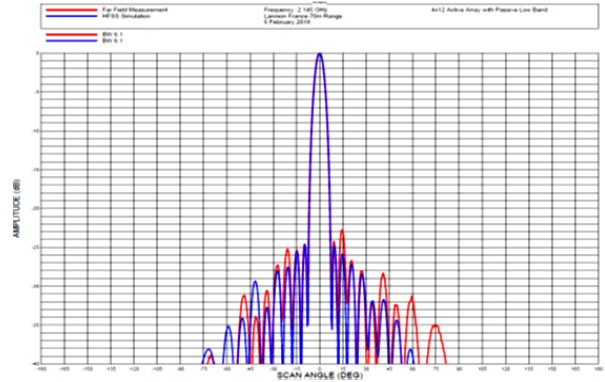}} 
\caption{Comparing bore-sight measured radiation patterns against simulation.}
\label{fig:beam}
\end{figure*}

In order to validate the performance of the coherent LO distribution and the proposed calibration scheme we measured the radiated beam patterns in both a near field anechoic chamber and far-field radiated testing range. The phase and magnitude coefficients applied to the mixed signal ICs correspond to both with and without amplitude taper and bore-sight beam. RF calibration is activated in a continuous sequential mode so as to access its impact on the radiated pattern. Near-field test results for narrow beam generated with and without taper are shown in Fig.~\ref{fig:beam}. The array under test corresponds to the system specified earlier for HDAAS. Hence the radiation pattern corresponds to a 12x4 array with 48 elements in one polarization. Both Azimuth and Elevation cuts of the 3D radiation pattern are shown. The red and white plots correspond to the high frequency structure simulator (HFSS) results corresponding to the exact models of antenna elements and spacing used. So these are ideal expected beam patterns from the array used. The blue and yellow lots are the measured results from the near-field measurements. It can be observed that there is excellent match over the main lobe. the side-lobes are about 1-3 dB off. More importantly the null depths have a mismatch of less than 5 dB. This corresponds to a RMS phase error of 5◦ and RMS magnitude error of 1 dB. The mismatch of the azimuth side-lobes beyond 50◦ is due to the limited range of the near field probe which could only traverse to the edge of the array. The radiated beam from far-field test is shown in Fig.~\ref{fig:beam:b}. This shows the azimuth cut of the measured radiation pattern. The blue plot corresponds to the expected beam pattern result from HFSS. The red plot corresponds to the measured result. There is excellent match over the main-lobe. There is less than 5dB of mismatch over the side-lobe levels especially closer to the main lobe. More important from MU-MIMO user separation perspective it can be observed that there is a good match over the null depths. It can also be observed that the sidelobe discrepancy observed in near-field azimuth cut beyond 50◦ is not found in far-field result which further validates the array performance. Based on these results it can be confirmed that the array phase center based on BDS LO distribution method is maintained across the array. The sequential calibration procedure does not impact the radiation pattern and keeps the radiation pattern within the expected phase and magnitude error limits even with internal temperature variation during operation.

Encouraged by these results several field trails have been conducted in various networks carrying commercial LTE traffic. The results of one of these trials have been shared in [9] which validates the fact that all the elements highly synchronized. HDAAS with full-connected hybrid beamforming architecture with 4 antenna ports is implemented with legacy Rel-8/10 FDD LTE system. In these deployments cell re-shaping by modifying the phase and magnitude coefficients of the mixed signal ICs was carried out to manage inter-cell interference and improve DL and UL SINR. In addition split-sector implementation with cell shape optimization was also carried out using the same array as described in~\cite{Erkip2018}. Drive tests as described in \cite{Vaghefi2018} were used to confirm the effect of beam optimization on the ground. These results would not have been possible without achieving the degree of coherency with LO distribution and active calibration. Based on this we can confidently state that the HDAAS can be re-used to implement a class-B Beamformed-CSI feedback based FD-MIMO system with fewer data converters and less digital processing. The same architecture to build each RF chain along with BDS based LO distribution and calibration can be extended to build a sub-array hybrid beamforming system supporting class-A non-precoded codebook based FD-MIMO system in FDD and reciprocity based transmission in TDD. Of course this will bear higher cost due to higher count of data converters and digital processing involved. Future massive MIMO systems that require scaling of RF chains can be enabled by the same techniques if the complexity and power consumed by the data converters and digital processing can be reduced.

\section{Conclusions} \label{conclusions}
Transforming any novel communications systems idea from concept to real world requires overcoming practical impairments. In the case of FD-MIMO proposed in 3GPP standards for cellular deployment this requires overcoming practical challenges imposed by RF impairments. In this paper we have focused on the challenges with overcoming multiplicative noise. 

A few observations can be summarized as follows:
i) If digitizing larger number of elements leads to more phase and magnitude errors then it is beneficial to limit the number of antenna ports with better coherency to achieve better capacity. It is better to implement sub-array hybrid beamforming systems with limited ports for TDD or a full-connected hybrid beamforming system for FDD with better coherency.  
ii) Beyond $10 \degree$ RMS phase error and $0.5 \mathrm{dB}$ magnitude error the performance of MF precoder with channel reciprocity and channel feedback based FD-MIMO performs closer to ZF based precoder with channel reciprocity. Hence, without high coherency capacity scaling expected with ZF precoder cannot be achieved. 
iii) Generating coherent LOs across RF chains is critical. SLO architecture with cost effective oscillators will require frequent re-calibration due to short term phase drift. PLL and CLO based methods are sensitive to internal temperature variations which requires careful design. Hence, while undertaking system design, stability of LOs is an important factor to consider to analyze performance vs. cost/implementation complexity tradeoff.
iv) Internal temperature variations is a major contributor to multiplicative noise. We discussed measurements taken from an active system on a tower top to characterize its effect on phase and magnitude response deviations. It is important for a systems engineer to understand stability of frequency response of the array design to temperature fluctuations and the effectiveness of calibration to compensate them in a timely manner with implications on spectral efficiency and additional overhead. This again will help a system designer to make realistic tradeoffs in terms of choice of array architecture and precoding. 

We discussed a novel architecture for coherent high frequency LO distribution and absolute calibration method that is applicable to both FDD and TDD that can be coherent with the timing reference generated by baseband unit. Various measurement results from lab, measured radiation patterns from near and far field and field trial results carrying commercial LTE network traffic prove the efficacy of these methods. 

\footnotesize
\bibliographystyle{IEEEtran}
\bibliography{main}

\begin{thebibliography}{10}
\providecommand{\url}[1]{#1}
\csname url@samestyle\endcsname
\providecommand{\newblock}{\relax}
\providecommand{\bibinfo}[2]{#2}
\providecommand{\BIBentrySTDinterwordspacing}{\spaceskip=0pt\relax}
\providecommand{\BIBentryALTinterwordstretchfactor}{4}
\providecommand{\BIBentryALTinterwordspacing}{\spaceskip=\fontdimen2\font plus
\BIBentryALTinterwordstretchfactor\fontdimen3\font minus
  \fontdimen4\font\relax}
\providecommand{\BIBforeignlanguage}[2]{{%
\expandafter\ifx\csname l@#1\endcsname\relax
\typeout{** WARNING: IEEEtran.bst: No hyphenation pattern has been}%
\typeout{** loaded for the language `#1'. Using the pattern for}%
\typeout{** the default language instead.}%
\else
\language=\csname l@#1\endcsname
\fi
#2}}
\providecommand{\BIBdecl}{\relax}
\BIBdecl

\bibitem{SamsungFDmimo2017}
H.~Ji, Y.~Kim, J.~Lee, E.~Onggosanusi, Y.~Nam, J.~Zhang, B.~Lee, and B.~Shim,
  ``Overview of full-dimension {MIMO} in {LTE}-advanced pro,'' \emph{IEEE
  Commun. Mag.}, vol.~55, no.~2, pp. 176--184, Feb 2017.

\bibitem{ShafiJsac20175G}
M.~Shafi, A.~F. Molisch, P.~J. Smith, T.~Haustein, P.~Zhu, P.~D. Silva,
  F.~Tufvesson, A.~Benjebbour, and G.~Wunder, ``{5G}: a tutorial overview of
  standards, trials, challenges, deployment, and practice,'' \emph{IEEE J. Sel.
  Areas Commun}, vol.~35, no.~6, pp. 1201--1221, 2017.

\bibitem{3gppAAs2015}
``Study on elevation beamforming/full-dimension {(FD) MIMO} for {LTE} (release
  13),'' 3GPP, Tech. Rep. TR 36.897, Jun 2015.

\bibitem{yang2013performance}
H.~{Yang} and T.~L. {Marzetta}, ``Performance of conjugate and zero-forcing
  beamforming in large-scale antenna systems,'' \emph{IEEE J. Sel. Areas
  Commun.}, vol.~31, no.~2, pp. 172--179, February 2013.

\bibitem{Ericsson2016}
``On the use of channel reciprocity for {NR},'' 3GPP, Ericsson, Tech. Rep. 3GPP
  TSG RAN WG1 Meeting 84, R1-163239, Apr 2016.

\bibitem{Nokia3GPP2017}
``On the channel reciprocity support for {CSI} acquisition,'' 3GPP, Nokia,
  Tech. Rep. 3GPP TSG RAN WG1 90bis, R1-1718706, Oct 2017.

\bibitem{Vook2018}
F.~W. {Vook}, W.~J. {Hillery}, E.~{Visotsky}, J.~{Tan}, X.~{Shao}, and
  M.~{Enescu}, ``System level performance characteristics of sub-{6GHz} massive
  {MIMO} deployments with the {3GPP} new radio,'' in \emph{Proc. IEEE VTC}, Aug
  2018, pp. 1--5.

\bibitem{Ericsson5GNRTestbed2018}
B.~{Halvarsson}, A.~{Simonsson}, A.~{Elgcrona}, R.~{Chana}, P.~{Machado}, and
  H.~{Asplund}, ``{5G NR} testbed 3.5 {GHz} coverage results,'' in \emph{Proc.
  IEEE VTC}, Jun 2018, pp. 1--5.

\bibitem{Vaghefi2018}
R.~M. Vaghefi, G.~Miranda, R.~Srirambhatla, G.~Marzin, C.~Ng, F.~Fayazbakhsh,
  S.~Tarigopula, R.~C. Palat, and M.~Banu, ``First commercial hybrid massive
  {MIMO} system for sub-6hz bands,'' in \emph{Proc. IEEE 5GWF}, Jul 2018, pp.
  357--362.

\bibitem{Anokiwave2017}
R.~{McMorrow}, D.~{Corman}, and A.~{Crofts}, ``All silicon mmw planar active
  antennas: The convergence of technology, applications, and architecture,'' in
  \emph{IEEE COMCAS}, Nov 2017, pp. 1--4.

\bibitem{Honcharenko2019}
W.~Honcharenko, ``Sub-6 {GHz} {mMIMO} base stations meet {5Gs} size and weight
  challenges,'' Feb 2019.

\bibitem{3GPP2017104}
``Multi-standard radio {(MSR)} base station {(BS)} radio transmission and
  reception,'' 3GPP, Tech. Rep. TS 37.104, Jun 2017.

\bibitem{3GPP2019}
``{NR}; base station {(BS)} radio transmission and reception,'' 3GPP, Tech.
  Rep. TS 38.104 V15.5.0, Mar 2019.

\bibitem{Zhong2012}
C.~Shepard, H.~Yu, N.~Anand, E.~Li, T.~Marzetta, R.~Yang, and L.~Zhong,
  ``Argos: Practical many-antenna base stations,'' in \emph{Proc. MobiCom},
  2012, pp. 53--64.

\bibitem{Tufvesson2014}
J.~Vieira, S.~Malkowsky, K.~Nieman, Z.~Miers, N.~Kundargi, L.~Liu, I.~Wong,
  V.~Öwall, O.~Edfors, and F.~Tufvesson, ``A flexible 100-antenna testbed for
  massive {MIMO},'' in \emph{Proc. IEEE GLOBECOM Workshops}, Dec 2014, pp.
  287--293.

\bibitem{HausteinPoCMamimo2016}
T.~Wirth, L.~Thiele, M.~Kurras, M.~Mehlhose, and T.~Haustein, ``Massive {MIMO}
  proof-of-concept: Emulations and hardware-field trials at 3.5 {GHz},'' in
  \emph{Proc. Asilomar}, Nov 2016, pp. 1793--1798.

\bibitem{Beach2018}
W.~B. Hasan, P.~Harris, A.~Doufexi, and M.~Beach, ``Real-time maximum spectral
  efficiency for massive mimo and its limits,'' \emph{IEEE Access}, vol.~6, pp.
  46\,122--46\,133, 2018.

\bibitem{Wei2017}
L.~{Dong}, H.~{Zhao}, Y.~{Chen}, D.~{Chen}, T.~{Wang}, L.~{Lu}, B.~{Zhang},
  L.~{Hu}, L.~{Gu}, B.~{Li}, H.~{Yang}, H.~{Shen}, T.~{Tian}, Z.~{Luo}, and
  K.~{Wei}, ``Introduction on {IMT-2020 5G} trials in {China},'' \emph{IEEE J.
  Sel. Areas Commun.}, vol.~35, no.~8, pp. 1849--1866, Aug 2017.

\bibitem{Kishiyama2017}
J.~{Wang}, A.~{Jin}, D.~{Shi}, L.~{Wang}, H.~{Shen}, D.~{Wu}, L.~{Hu}, L.~{Gu},
  L.~{Lu}, Y.~{Chen}, J.~{Wang}, Y.~{Saito}, A.~{Benjebbour}, and
  Y.~{Kishiyama}, ``Spectral efficiency improvement with {5G} technologies:
  Results from field tests,'' \emph{IEEE J. Sel. Areas Commun.}, vol.~35,
  no.~8, pp. 1867--1875, Aug 2017.

\bibitem{Okumura2017}
K.~Yamazaki, T.~Sato, Y.~Maruta, T.~Okuyama, J.~Mashino, S.~Suyama, and
  Y.~Okumura, ``{DL MU-MIMO} field trial with {5G} low {SHF} band massive
  {MIMO} antenna,'' in \emph{Proc. IEEE VTC}, Jun 2017, pp. 1--5.

\bibitem{Okumura2018}
Y.~Maruta, K.~Yamazaki, K.~Izui, T.~Okuyama, J.~Mashino, S.~Suyama,
  K.~Nakayasu, T.~Sato, and Y.~Okumura, ``Outdoor {DL MU-MIMO} and inter access
  point coordination performance of low-{SHF}-band {C-RAN} massive {MIMO}
  system for {5G},'' in \emph{Proc. IEEE VTC}, Jun 2018, pp. 1--5.

\bibitem{Debbah2015}
E.~Björnson, M.~Matthaiou, and M.~Debbah, ``Massive {MIMO} with non-ideal
  arbitrary arrays: Hardware scaling laws and circuit-aware design,''
  \emph{IEEE Trans. Wireless Commun.}, vol.~14, no.~8, pp. 4353--4368, Aug
  2015.

\bibitem{Ratnarajah2017}
A.~Papazafeiropoulos and T.~Ratnarajah, ``Toward a realistic assessment of
  multiple antenna {HCNs}: Residual additive transceiver hardware impairments
  and channel aging,'' \emph{IEEE Trans. Veh. Technol.}, vol.~66, no.~10, pp.
  9061--9073, Oct 2017.

\bibitem{Kaltenberger2018}
X.~{Jiang} and F.~{Kaltenberger}, ``Channel reciprocity calibration in {TDD}
  hybrid beamforming massive {MIMO} systems,'' \emph{IEEE J. Sel. Topics Signal
  Process.}, vol.~12, no.~3, pp. 422--431, Jun 2018.

\bibitem{SamsungJsac2017FDMimoProto}
G.~Xu, Y.~Li, J.~Yuan, R.~Monroe, S.~Rajagopal, S.~Ramakrishna, Y.~Nam,
  J.~Seol, J.~Kim, M.~Gul, A.~Aziz, and J.~Zhang, ``Full dimension {MIMO}
  ({FD-MIMO}): Demonstrating commercial feasibility,'' \emph{IEEE J. Sel. Areas
  Commun}, vol.~35, no.~8, pp. 1876--1886, Aug 2017.

\bibitem{Tufvesson2017}
J.~{Vieira}, F.~{Rusek}, O.~{Edfors}, S.~{Malkowsky}, L.~{Liu}, and
  F.~{Tufvesson}, ``Reciprocity calibration for massive mimo: Proposal,
  modeling, and validation,'' \emph{IEEE Trans. Wireless Commun.}, vol.~16,
  no.~5, pp. 3042--3056, May 2017.

\bibitem{Larsson2015}
A.~{Pitarokoilis}, S.~K. {Mohammed}, and E.~G. {Larsson}, ``Uplink performance
  of time-reversal mrc in massive mimo systems subject to phase noise,''
  \emph{IEEE Trans. Wireless Commun.}, vol.~14, no.~2, pp. 711--723, Feb 2015.

\bibitem{Krishnan2016}
R.~Krishnan, M.~R. Khanzadi, N.~Krishnan, Y.~Wu, A.~G. i~Amat, T.~Eriksson, and
  R.~Schober, ``Linear massive {MIMO} precoders in the presence of phase
  noise—a large-scale analysis,'' \emph{IEEE Trans. Veh. Technol.}, vol.~65,
  no.~5, pp. 3057--3071, May 2016.

\bibitem{PuglielliBwrcPnBF2016}
A.~Puglielli, G.~LaCaille, A.~M. Niknejad, G.~Wright, B.~Nikolić, and E.~Alon,
  ``Phase noise scaling and tracking in {OFDM} multi-user beamforming arrays,''
  in \emph{Proc. IEEE ICC}, May 2016, pp. 1--6.

\bibitem{Dai2015}
W.~{Zhang}, H.~{Ren}, C.~{Pan}, M.~{Chen}, R.~C. {de Lamare}, B.~{Du}, and
  J.~{Dai}, ``Large-scale antenna systems with {UL/DL} hardware mismatch:
  achievable rates analysis and calibration,'' \emph{IEEE Trans. Commun.},
  vol.~63, no.~4, pp. 1216--1229, Apr 2015.

\bibitem{Luo2016}
X.~{Luo}, ``Multiuser massive {MIMO} performance with calibration errors,''
  \emph{IEEE Trans. Wireless Commun.}, vol.~15, no.~7, pp. 4521--4534, Jul
  2016.

\bibitem{Tafazolli2017}
D.~{Mi}, M.~{Dianati}, L.~{Zhang}, S.~{Muhaidat}, and R.~{Tafazolli}, ``Massive
  mimo performance with imperfect channel reciprocity and channel estimation
  error,'' \emph{IEEE Trans. Commun.}, vol.~65, no.~9, pp. 3734--3749, Sep.
  2017.

\bibitem{3gpp:3dchannel}
``Study on {3D} channel model for {LTE},'' no. {3GPP TR 36.873}, 2018.

\bibitem{Madhow2012}
D.~R. {Brown}, P.~{Bidigare}, S.~{Dasgupta}, and U.~{Madhow},
  ``Receiver-coordinated zero-forcing distributed transmit nullforming,'' in
  \emph{Proc. IEEE SSP}, Aug 2012, pp. 269--272.

\bibitem{Willwerth1995}
H.~M. {Aumann} and F.~G. {Willwerth}, ``Phased array calibrations using
  measured element patterns,'' in \emph{Proc. IEEE APS}, vol.~2, Jun 1995, pp.
  918--921.

\bibitem{Kuehnke2001}
L.~{Kuehnke}, ``Phased array calibration procedures based on measured element
  patterns,'' in \emph{IEE APC}, vol.~2, Apr 2001, pp. 660--663.

\bibitem{Galton2019}
I.~{Galton} and C.~{Weltin-Wu}, ``Understanding phase error and jitter:
  Definitions, implications, simulations, and measurement,'' \emph{IEEE Trans.
  Circuits Syst. I, Reg. Papers}, vol.~66, no.~1, pp. 1--19, Jan 2019.

\bibitem{FetweissTcom2007}
D.~{Petrovic}, W.~{Rave}, and G.~{Fettweis}, ``Effects of phase noise on {OFDM}
  systems with and without {PLL}: Characterization and compensation,''
  \emph{IEEE Trans. Commun.}, vol.~55, no.~8, pp. 1607--1616, Aug 2007.

\bibitem{Rakon2009}
G.~Trudgen, ``Variance as applied to crystal oscillator,'' Rakon UK Ltd, Tech.
  Rep., 2009.

\bibitem{Kumar2014}
A.~{Kumar}, P.~A. {Koch}, H.~E. {Baidoo-Williams}, R.~{Mudumbai}, and
  S.~{Dasgupta}, ``An empirical study of the statistics of phase drift of
  off-the-shelf oscillators for distributed {MIMO} applications,'' in
  \emph{Proc. IEEE DYSPAN}, Apr 2014, pp. 350--353.

\bibitem{Brown2017}
J.~{McNeill}, S.~{Razavi}, K.~{Vedula}, and D.~{Richard Brown}, ``Experimental
  characterization and modeling of low-cost oscillators for improved carrier
  phase synchronization,'' in \emph{Proc. IEEE I2MTC}, May 2017, pp. 1--6.

\bibitem{Minihold2013}
O.~Werther and R.~Minihold, ``{LTE} system specifications and their impact on
  {RF} \& base band circuits,'' Rohde \& Schwarz, Tech. Rep., Apr 2013.

\bibitem{Zucca2005}
C.~{Zucca} and P.~{Tavella}, ``The clock model and its relationship with the
  allan and related variances,'' \emph{IEEE Trans. Ultrason., Ferroelectr.,
  Freq. Control}, vol.~52, no.~2, pp. 289--296, Feb 2005.

\bibitem{Vig1992}
J.~Vig, ``Introduction to quartz frequency standards,'' Army Research
  Laboratory, Electronics and Power Sources Directorate, Tech. Rep., 1992.

\bibitem{Mehrpouyan2012}
H.~Mehrpouyan, A.~A. Nasir, S.~D. Blostein, T.~Eriksson, G.~K. Karagiannidis,
  and T.~Svensson, ``Joint estimation of channel and oscillator phase noise in
  {MIMO} systems,'' \emph{IEEE Trans. Signal Process.}, vol.~60, no.~9, pp.
  4790--4807, Sep 2012.

\bibitem{Keysight2014}
``Signal source solutions for coherent and phase stable multi-channel
  systems,'' Keysight Technologies, Tech. Rep., July 2014.

\bibitem{ADIRefDist2018}
P.~Delos, ``System-level lo phase noise model for phased arrays with
  distributed phase-locked loops,'' Analog Devices, Inc., Tech. Rep., 2018.

\bibitem{TIJesd2015}
M.~Guibord, ``{JESD204B} multi-device synchronization: Breaking down the
  requirements,'' Texas Instruments, Inc., Tech. Rep., 2015.

\bibitem{ADIJesd204B}
A.~Oz and K.~Peker, ``{JESD204B} survival guide practical {JESD204B} technical
  information, tips, and advice from the world’s data converter market share
  leader,'' Analog Devices, Inc., Tech. Rep., 2013.

\bibitem{Hall2014}
D.~Hall, A.~Hinde, and Y.~Jia, ``Multi-channel phase-coherent {RF} measurement
  system architectures and performance considerations,'' in \emph{Proc. IEEE
  MILCOM}, Oct 2014, pp. 1318--1323.

\bibitem{RhodeSchwarz2016}
``Generating multiple phase coherent signals – aligned in phase and time,''
  Rohde \& Schwarz Inc., Tech. Rep., Sep 2016.

\bibitem{ADIClkSkew2019}
C.~Pearson, ``Clock skew in large multi-{GHz} clock trees,'' Analog Devices,
  Inc., Tech. Rep., Jan 2019.

\bibitem{Beach1998}
G.~{Tsoulos}, J.~{McGeehan}, and M.~{Beach}, ``Space division multiple access
  ({SDMA}) field trials. 2. calibration and linearity issues,'' \emph{IEE
  Proc., Radar, Sonar and Navig.}, vol. 145, no.~1, pp. 79--84, Feb 1998.

\bibitem{Brauner2003}
T.~{Brauner}, R.~{Kung}, R.~{Vogt}, and W.~{Bachtold}, ``5-6 {GHz} low-noise
  active antenna array for multi-dimensional channel-sounding,'' in \emph{Proc
  SBMO/IEEE MTT-S}, vol.~1, Sep 2003, pp. 297--301.

\bibitem{Aghvami2007}
N.~Tyler, B.~Allen, and A.~H. Aghvami, ``Calibration of smart antenna systems:
  measurements and results,'' \emph{IET Microwaves, Antennas Propagation},
  vol.~1, no.~3, pp. 629--638, Jun 2007.

\bibitem{KaltenbergerFmwkCal2018}
X.~{Jiang}, A.~{Decurninge}, K.~{Gopala}, F.~{Kaltenberger}, M.~{Guillaud},
  D.~{Slock}, and L.~{Deneire}, ``A framework for over-the-air reciprocity
  calibration for {TDD} massive {MIMO} systems,'' \emph{IEEE Trans. Wireless
  Commun.}, vol.~17, no.~9, pp. 5975--5990, Sep 2018.

\bibitem{banu2012method}
M.~Banu and V.~Prodanov, ``Method and system for multi-point signal generation
  with phase synchronized local carriers,'' Sep 2012, {US} Patent 8,259,884.

\bibitem{Banu2006VlsiClk}
V.~Prodanov and M.~Banu, ``{GHz} serial passive clock distribution in {VLSI}
  using bidirectional signaling,'' in \emph{Proc. IEEE CICC}, Sep 2006, pp.
  285--288.

\bibitem{Erkip2018}
S.~{Shahsavari}, S.~A. {Hosseini}, C.~{Ng}, and E.~{Erkip}, ``Adaptive hybrid
  beamforming with massive phased arrays in macro-cellular networks,'' in
  \emph{IEEE 5GWF}, Jul 2018, pp. 221--226.

\end{thebibliography}

\end{document}